\documentclass{aastex}
\usepackage{emulateapj5}
\usepackage{epsf}
\usepackage{onecolfloat}

\def\LCDM{$\Lambda$CDM}
\newcommand{\lya}{Ly-$\alpha$ }

\begin{document}
\submitted {draft version; \today}

\shortauthors{TASITSIOMI}
\shorttitle{\lya radiative transfer in  cosmological simulations}
\twocolumn[%
\title{\lya Radiative Transfer in  Cosmological  Simulations and  application to 
a $\lowercase{z}\simeq 8$ \lya Emitter}

\author{Argyro Tasitsiomi\altaffilmark{1,2}} 
\vspace{2mm}
\begin{abstract}
We develop a  \lya radiative transfer (RT)  Monte Carlo code for cosmological simulations. 
High resolution,  along with appropriately treated cooling,  can result in  
 simulated environments with very high optical depths.
Thus, solving the \lya RT problem in cosmological simulations can  take an unrealistically long time. 
For this reason,  we develop  methods to speed up the \lya RT.
With these accelerating methods, along with the parallelization of the code, we make the problem of \lya RT in the complex environments of
cosmological simulations tractable. 
We  test the RT code against simple \lya emitter models, and then we apply it to the brightest \lya emitter of a
gasdynamics+N-body  Adaptive Refinement Tree (ART) simulation at $z \simeq 8$.
We find that recombination  rather than cooling radiation \lya photons is the dominant 
contribution  to the intrinsic \lya luminosity of the emitter, which is  $\simeq 4.8 \times 10^{43}$ ergs/s.
The size of the  emitter is pretty small, making it  unresolved for currently available instruments. 
Its spectrum  before adding the \lya Gunn-Peterson absorption (GP)  resembles that of static media, 
despite some net inward radial peculiar motion. This is because for such high optical depths as those in ART simulations, velocities
of order some hundreds km/s are not important.
We add the GP in two ways. First we assume  no damping wing, 
corresponding  to the situation where the emitter lies within the HII region of a very bright quasar, and second we allow for  the damping wing.
Including the damping wing
 leads to a maximum line brightness suppression by
roughly a factor of $\sim 62$. 
The line fluxes, even though quite faint for current ground-based telescopes, should be within reach for JWST.

\end{abstract}


]

\altaffiltext{1}{Dept. of Astronomy and Astrophysics,
                Kavli Institute for Cosmological Physics,
                The University of Chicago, Chicago, IL 60637}
\altaffiltext{2}{Current address: Dept. of Astrophysical Sciences, 
                 Princeton University, Peyton Hall-Ivy Lane,
                 Princeton, NJ 08544; {\tt iro@astro.princeton.edu}}
\section{Introduction}
\label{sec:intro}
Since the classic paper by \cite{partridge_peebles67},  intense observational efforts have focused 
on the search for \lya emitters at high redshifts. 
Although most of the early attempts ended in negative results before the mid 1990s, recent observational advances enabled us to identify star forming galaxies
 at ever   
increasing redshifts. Currently, 
several observational projects, such as LALA \cite[e.g.,][]{rhoads_etal03}, 
CADIS \cite[e.g.,][]{maier02}, the Subaru Deep Field Project \cite[e.g.,][]{taniguchi_etal05}, etc.,
spectroscopic  surveys that use lensing magnification from clusters \citep[e.g.,][]{santos_etal03}, surveys that combine 
Subaru  \citep[e.g.,][]{hu_etal04} or HST/ACS/NICMOS  imaging \citep[e.g.,][for the GOODS survey]{stanway_etal04, dickinson_etal04} with Keck spectroscopy,  etc., 
focus on  finding high-z starforming galaxies. 
Surveys currently reach up to $z\simeq  7-8$  \cite[e.g.,  see][for recent results from the NICMOS observations of the HUDF]{bouwens_etal04} and will likely 
reach higher redshifts in the coming years (e.g., via  JWST).  

The hydrogen \lya line is a very promising way to probe the high-redshift universe. Besides yielding redshifts, the shape, equivalent width and offset of the \lya line 
from other emission/absorption lines potentially convey valuable information about the geometry, kinematics, and underlying stellar population of the host galaxy.
Furthermore, after escaping the environment of the host galaxy, \lya photons are scattered in the surrounding IGM. The presence or absence of observed
\lya emission can be used to place constraints on the state of the IGM, useful in constraining for example the epoch and topology 
of reionization. Because of the
numerous factors that contribute to the final \lya emission, the interpretation of such features can be very complex.  
To use all the currently available and future observations in the most effective way possible we need to improve  
our theoretical understanding of \lya emission from high-redshift objects. 
To this end we develop a general \lya radiative transfer (RT) scheme for cosmological simulations. 
As an example,  we apply  the RT scheme to 
gasdynamics+N-body Adaptive Refinement Tree \citep[ART;][]{kravtsov03} 
simulations of galaxy formation.

There are quite a few studies of \lya emission from high-z objects \cite[e.g.,][]{gould_weinberg96,haiman_spaans99,loeb_rybicki99,ahn_etal01,ahn_etal02b,zheng_miralda-escude02,
santos04,dijkstra_etal05a,dijkstra_etal05b}. The problem these studies address is highly complex and has many unknowns. Inevitably,  most of these studies 
had to make at  some point some simplifying 
assumptions. Usually,  a high degree of symmetry for the emitting source, and its density, temperature, and velocity field is assumed. 
The same is the case with respect to the processes that are responsible for the production of \lya photons.
On the other hand,  cosmological simulations hopefully capture most of the basic elements, lifting thus practical constraints that existed in these previous studies.

There is a small number of related studies using 
cosmological simulations \citep{fardal_etal01,furlanetto_etal03,barton_etal04,furlanetto_etal05,cantalupo_etal05,ledelliou_etal05a,ledelliou_etal05b}. 
Some of these simulations are  lacking crucial processes  such as  radiative cooling of the gas and consistent RT, 
the
various sources of \lya photons,
and/or sufficient resolution in order to resolve the clumpiness of the gas.  
Furthermore,  most of these studies
do not perform \lya RT, but rather they assume that the observer sees whatever is being emitted initially, 
simply modified by $e^{-\tau}$ with $\tau$ the optical depth for \lya scattering due to neutral hydrogen between the emission
point and the observer. 
Namely, in most cases \lya spectra from simulations are treated as {\it absorption} spectra
when, in reality, they are {\it scattering} spectra \cite[see, e.g.,][]{gnedin_prada04}.  For gas well outside  the source of emission this is an
appropriate approximation since scattering off the direction of viewing removes the photons that could be observed and thus appears as effective
absorption.  This is no longer true for the source of emission itself, since photons that were originally emitted in directions different from
the direction of observation may scatter into this direction. 

It is important that the difficulty of implementing a \lya RT scheme for cosmological simulations become clear.
The classical problem of resonance RT, 
 relevant to  a wide range of applications
 from planetary atmospheres to  accretion disks, has been quite extensively studied in the 
 literature \cite[e.g.,][]{zanstra49,hummer62,auer68,avery_house68,adams72,
harrington73,neufeld90,ahn_etal01,ahn_etal02b}.
However, analytical solutions derived in the past are applicable only to certain specific conditions.
On the other hand, the slow convergence of the numerical techniques used limited the
numerical studies at optical thicknesses that are relatively low compared to those encountered in high-redshift galaxies (and cosmological simulations 
of high-redshift galaxies, as we will show).
Thus, unlike 
previous studies, most of which  focused on the classical problem of resonant RT in a 
semi-infinite slab, in cosmological simulations one 
has to solve simultaneously thousands or even millions of these problems.\footnote{For example, in the case of  Adaptive Mesh Refinement (AMR) codes, 
each time a photon enters a simulation cell one has the equivalent of a new slab RT problem.}Furthermore, having in mind
existing and future cosmological simulations that can achieve  sufficiently high resolution to resolve the gas clumpiness and that treat cooling appropriately,  
we anticipate column densities that are orders of magnitude higher than those
found in lower resolution simulations without cooling. In this case we need a RT algorithm much faster than
the more standard direct Monte Carlo approach [which, however, is our starting point] of previous studies.   
Thus,  we must develop  RT acceleration methods that, along with the highly parallel nature of the RT problem that enables us
to make use of many parallel machines, can make the \lya RT problem tractable.

The paper is organized as follows. In \S~\ref{sec:rt} we discuss the RT scheme.
More specifically, in \S~\ref{montecarlo} we present the basic Monte Carlo algorithm, 
in \S~\ref{sec:test} we present tests of the basic algorithm, in \S~\ref{sec:accel} we discuss
the acceleration methods we use  to speed up the RT, and in \S~\ref{images_spectra} we present the method images and spectra are constructed.
In \S~\ref{app_sims} we discuss in detail an application of the \lya RT code to ART simulations.
More specifically, in \S~\ref{sec:sims} we briefly give some information
about the ART simulations. In \S~\ref{intrinsic_emission} we discuss the
intrinsic \lya emission of the specific  simulated \lya emitter we focus on.  
In \S~\ref{beforert} we present results on  the emitter before RT.  
In \S~\ref{afterrt} we discuss results after performing the RT, and with/without the Gunn-Peterson (GP) absorption,
as well as with/without the red damping wing of the GP absorption.
In \S~\ref{sec:conclusions} we discuss and summarize our results and
conclusions. 
\section{The \lya Radiative Transfer}
\label{sec:rt}
\subsection{The basic Monte Carlo code}
\label{montecarlo}
The following discussion  assumes in various places a cell structure for the simulation outputs, as is inherently the case in AMR codes.
However, the \lya RT code we discuss  is applicable to outputs from all kinds of cosmological simulation codes, since 
one can always create an effective mesh by interpolating the values of the various physical parameters. The 
size of the mesh cell can be motivated by  resolution related
 scales (e.g., the softening scale, or
larger if convergence tests with respect to the \lya RT justify a larger scale).
Thus, in what follows we  refer to simulation cells either the direct output of the cosmological simulation code has a cell structure or
not.    

The initial emission characteristics (simulation cell, frequency, etc.)  of each
photon depend on the specific physical conditions, thus  we defer this discussion for \S \ref{app_sims} where an
application to a \lya emitter produced in ART cosmological simulations is presented.
After determining the initial characteristics for each photon, we follow a series of scatterings up to a certain scale
where the detailed RT stops. This scale is to be determined via a convergence study.
In this subsection  we describe the basic steps of the algorithm.

\subsubsection{Propagating the photon}
For every scattering   we generate the optical depth, which determines the spatial displacement of the photon,
  by sampling the probability
distribution function $e^{-\tau}$ 
\begin{equation}
\tau=-\ln(R) \, ,
\label{taur}
\end{equation}
with $R$ a uniformly distributed random number.
This optical depth is equal to
\begin{equation} 
\tau=\int\limits_{0}^l \int\limits_{-\infty}^{\infty} d\tilde{l}du_{p} \sigma_{L}(\nu(1-u_{p}/c)) \sqrt{\frac{m_{p}}{2 \pi k_{B} T}}
n_{HI}\exp{\left({-\frac{m_{p} u_{p}^2}{2k_{B}T}}\right)} \, , 
\label{tau}
\end{equation}
with $n_{HI}$ the number density of neutral hydrogen. The function  $\sigma_{L}$ is the scattering cross section
of \lya photons as a function of frequency, defined in the rest frame of the hydrogen atom as
\begin{equation}
\sigma_{L}(\nu) = f_{12} \frac{\pi e^2}{m_{e} c} \frac{\Delta \nu_{L}/2 \pi}{(\nu-\nu_{0})^2+ (\Delta \nu_{L}/2)^2} \, ,
\label{sigma}
\end{equation}
where $f_{12}=0.4162$ is the \lya oscillator strength, $\nu_{0}=2.466 \times 10^{15}$ Hz is the line center frequency,
$\Delta \nu_{L}=9.936 \times 10^7$ Hz is the natural width of the line, and other symbols have their usual meaning.
In equation (\ref{tau}) the fact that the photons are encountering atoms with a Maxwellian distribution of thermal
velocities has been taken into account.
Integrating over  the  distribution of velocities, the resulting cross section  in the observer's frame is
\begin{equation}
\label{voigt}
\sigma(x)= f_{12} \frac{\sqrt{\pi} e^2}{m_{e} c \Delta \nu_{D}} H(\alpha,x)
\end{equation}
where 
\begin{equation}
H(\alpha,x)=\frac{\alpha}{\pi} \int_{-\infty}^{\infty} \frac{e^{-y^2}}{(x-y)^2 +\alpha^2} dy
\end{equation}
is the Voigt function, $x=(\nu-\nu_{0})/\Delta \nu_{D}$ is the relative frequency of the incident photon
in the observer's frame with  $\Delta \nu_{D}= \sqrt{2 k_{B} T/(m_{p} c^{2})} \nu_{0}$ the Doppler
width, and $\alpha=\Delta \nu_{L}/2 \Delta \nu_{D}$ with $\Delta \nu_{L}$ the natural line width. 
Assuming that $\sigma$ is independent of $\tilde{l}$, the optical depth  is given by
\begin{equation}
\tau=n_{HI} \sigma(x) l \, .
\label{tau2}
\end{equation}
When applied to  cosmological  simulations, equation~(\ref{tau2}) is substituted by a sum
of terms similar to the r.h.s. This sum is  over the different cells (=different physical conditions such as neutral hydrogen density, temperature, etc.) 
that the photon crosses until it reaches $\tau$ and gets scattered. 

For the Voigt function we 
use the following analytic fit, which is a good approximation to better than $1 \%$ for temperatures
$T> 2$K (N. Gnedin, personal communication)
\begin{eqnarray}
\nonumber
V(\alpha,\nu) & \equiv & \frac{1}{\sqrt{\pi} \Delta \nu_{D}} H(\alpha,x)= \frac{1}{\Delta \nu_{D}} \phi(x) \\ 
& = & \frac{1}{\Delta \nu_{D}} \left[ q + \frac{e^{-\tilde{x}}}{1.77245385} \right]
\end{eqnarray}
where $\tilde{x}=x^2$, and $q=0$ if $z=(\tilde{x}-0.855)/(\tilde{x}+3.42) \le 0$ and 
\begin{eqnarray}
\nonumber
q & = & z\left(1+\frac{21}{\tilde{x}}\right)\frac{\alpha}{\pi(\tilde{x}+1)} \\ 
& \times & \left\{0.1117 +z\left[4.421+z(-9.207+5.674z)\right]\right\}
\end{eqnarray}
if $z>0$. The definition in terms of the function $\phi(x)$ is also given since the latter has been used in many previous studies, and
we also use it in what follows. 
If in addition to the thermal motion of the atoms there is bulk motion, such as peculiar or Hubble flow velocities, 
in equation (\ref{voigt}) we  use
$x_{f}=x-(v_{fz}/c)\nu_{0}/ \Delta \nu_{D}$, where $v_{fz}$ is the component of the fluid bulk  velocity along the
direction of the incident photon.

In equation (\ref{tau}) the cross section $\sigma$ becomes $\tilde{l}$-dependent when Hubble expansion is taken into account. 
In this case
the equation is an integral and does not reduce to the simple algebraic equation~(\ref{tau2}). 
To propagate the photon one must solve for
the step which is the upper limit of the integral.  
In the simple examples discussed in \S \ref{sec:test}, 
things are relatively simple even when the Hubble expansion is included, since
in these cases there is homogeneity and isothermality and no sum over cells is required. In those cases, Hubble expansion is included as follows:
1.) we make a first guess for $l$ using the Hubble velocity at the current point, 2.) we use as a step for the photon a certain fraction of $l$,
3.) for
a specified tolerance  with which we want to achieve $\tau$, we refine the step as necessary. 
Note that the simple tests of the code presented in \S \ref{sec:test} do not include peculiar motions.  
In the actual simulations the peculiar velocities rather than the Hubble 
expansion are  dominant on the relevant scales (e.g., for the emitter we focus on, the
mean radial component of the peculiar motion dominates over the Hubble expansion up to about
$80$ physical kpc). In the detailed RT which we perform within such distances, we approximate the subdominant 
Hubble expansion velocity within a certain cell by the
expansion velocity that corresponds to the center of that cell.  This is calculated to have a negligible effect on the
results. 

The $n=2$ state of atomic hydrogen consists of the $2S_{1/2}, 2P_{1/2}$ and $2P_{3/2}$ substates,  whereas the
$n=1$ state consists of $1S_{1/2}$. 
According to the electric dipole selection 
rules, the allowed transitions are $2P_{1/2}$ to $1S_{1/2}$ and $2P_{3/2}$ to $1S_{1/2}$, whereas $2S_{1/2}$  corresponds
to destruction of the initially absorbed \lya photon, since this state de-excites through 
the emission of two continuum photons.
The multiplicity of each of these states is $2J+1$.  Thus the probabilities for the $2P$ states the atom
can be found in when absorbing the \lya photons  $2P_{1/2}:2P_{3/2}$ are
$1:2$. 
Collisions can potentially 
cause the $2P\rightarrow 2S$ transition in which case the photon gets destroyed.
A similar destruction effect can be caused by the existence of dust. Both these destruction mechanisms are
briefly discussed in the context of the ART simulations in \S \ref{sec:collisions} and \S \ref{sec:dust}, respectively. 
Considering the $2P_{1/2}$ and $2P_{3/2}$ cases separately, one would have to modify both the Voigt function
and the velocity distribution of the scattering atom discussed in the next section \citep[see, e.g., ][]{ahn_etal01}.
However, the level splitting between the two $2P$ states is small, just 10 GHz. This corresponds to a velocity width of $\sim 1$ km/s, much
smaller than the width due to thermal velocities in media with roughly $T>100$ K. In addition, even for lower temperatures, this level splitting
is still small for high optical depths. In our case, the thermal, peculiar, and Hubble velocities are all more important than the splitting, and 
combined with the
fact that we have high optical depths,  we do not make the distinction between the two sublevels.	  
As discussed below, however, the different fine structure levels are taken into account when choosing scattering  phase functions, important for
polarization calculations that we will present in a future paper.

\subsubsection{The scattering}
\label{scattering}
After determining the point in space where the photon will be  scattered  next, we choose the thermal velocity components
of the scattering atom. In the two directions perpendicular to the direction of the incident photon the components are drawn
from a (1-D)  Gaussian distribution  with dispersion equal to $\sqrt{\frac{k_{B} T}{m_{p}}}$. 
The component $u_{p}$ of the thermal velocity of the atom  along the direction
of the incident photon is drawn from the  distribution
\begin{equation}
f(v_{p})=\frac{a}{\pi} \frac{e^{-v_{p}^{2}}}{(x-v_{p})^2+a^2} H^{-1}(a,x)\, ,
\label{distr}
\end{equation}
with $v_{p}=u_{p} (m_{p}/ 2kT)^{1/2}$.
To draw numbers that follow this distribution we use the method of \citet{zheng_miralda-escude02}.

After each scattering we need to assign a new frequency (in the observer's frame) and direction to the photon.
To this end we perform a Lorentz transformation of the frequency and direction of the incident photon from the observer to
the atom rest frame, using the velocity of the atom  chosen as  described previously.
 
Although the code ignores the level splitting with respect to the scattering cross section and the velocity distribution, it takes into
account  the different phase distributions for core versus wing scatterings, as well as for $2P_{1/2}$ versus
$2P_{3/2}$ scatterings.
For resonant scattering, it is the angular momenta of the three states involved and the multipole order of the emitted
radiation that determines the scattering phase function.
\citet{hamilton40}  found that  the transition from $2P_{1/2}$  gives totally unpolarized photons 
and is characterized by an
isotropic angular distribution function, whereas that from the $2P_{3/2}$ state  corresponds to a maximum degree of polarization of 3/7 for a
$90^{\circ}$ scattering \citep[also see][]{chandrasekhar}.
More specifically, the scattering phase function for dipole transition can be written as \citep{hamilton40}
\begin{equation}
W(\theta) \propto 1+\frac{R}{Q} \cos^{2}\theta
\end{equation}   
with $R/Q$ the degree of polarization for a $90^{\circ}$ scattering and equal to
\begin{equation}
R/Q=\frac{(J+1)(2J+3)}{26J^{2}-15J-1} 
\end{equation}
for the $2P_{3/2} \rightarrow 1S_{1/2}$ transition since $\Delta J=-1,\Delta j=1, J=3/2$ according to Hamilton's conventions, and 
\begin{equation}
R/Q=\frac{(2J-1)(2J+3)}{12J^{2}+12J+1}
\end{equation}
for the $2P_{1/2} \rightarrow 1S_{1/2}$ transition with $\Delta J=\Delta j=0, J=1/2$. In 
both equations, J is the total angular momentum at the excited 
($n=2$) state. Thus,  $W(\theta)$ is constant (isotropic) for $2P_{1/2}$ as the excited state, whereas it equals 
\begin{equation}
W(\theta) \propto 1+ 3/7 \cos^{2}\theta
\end{equation}
with maximum polarization degree of $3/7$ at a $90^{\circ}$ scattering.

On the other hand, Stenflo (1980) showed that at high frequency shifts (i.e., at the line wings) quantum mechanical interference between the two
lines acts in such a way as to give a scattering behavior identical to that of a classical oscillator, namely
pure Rayleigh scattering.  Then the direction follows a dipole angular distribution
with Rayleigh polarization 100$\%$
at $90^0$ scattering, namely 
\begin{equation}
W(\theta) \propto 1+\cos^{2}\theta \, .
\end{equation}

Lastly,  
the frequency of the photon before and after scattering in the rest frame of the atom differs only by the recoil effect. Hence,
\begin{equation}
\tilde{\nu}=\frac{\nu}{ 1 + \frac{h \nu}{m_{p} c^{2}} (1- \cos\theta)}
\end{equation}
where $\nu, \tilde{\nu}$ are the frequency of the incident and scattered photon in the atom rest frame, respectively, the latter modified
due to the recoil effect. This effect is negligible for the environments produced in the simulations.

After determining the new direction and frequency of the scattered photon in the atom's rest frame we transform back to the observer's frame, and repeat the
whole scattering procedure.
\subsection{Testing the basic scheme}
\label{sec:test}
Here  we present some of the tests of the RT code we performed against analytical solutions, as well as other numerical
results  that exist in the literature.
In addition to showing the good performance of the code, these tests are presented here as  relevant to either \lya emitters and/or the way we accelerate the
code when applied in cosmological simulations (see \S \ref{sec:accel}).  

\begin{figure*}[thb]
\centerline{{\epsfxsize=3.5truein \epsffile{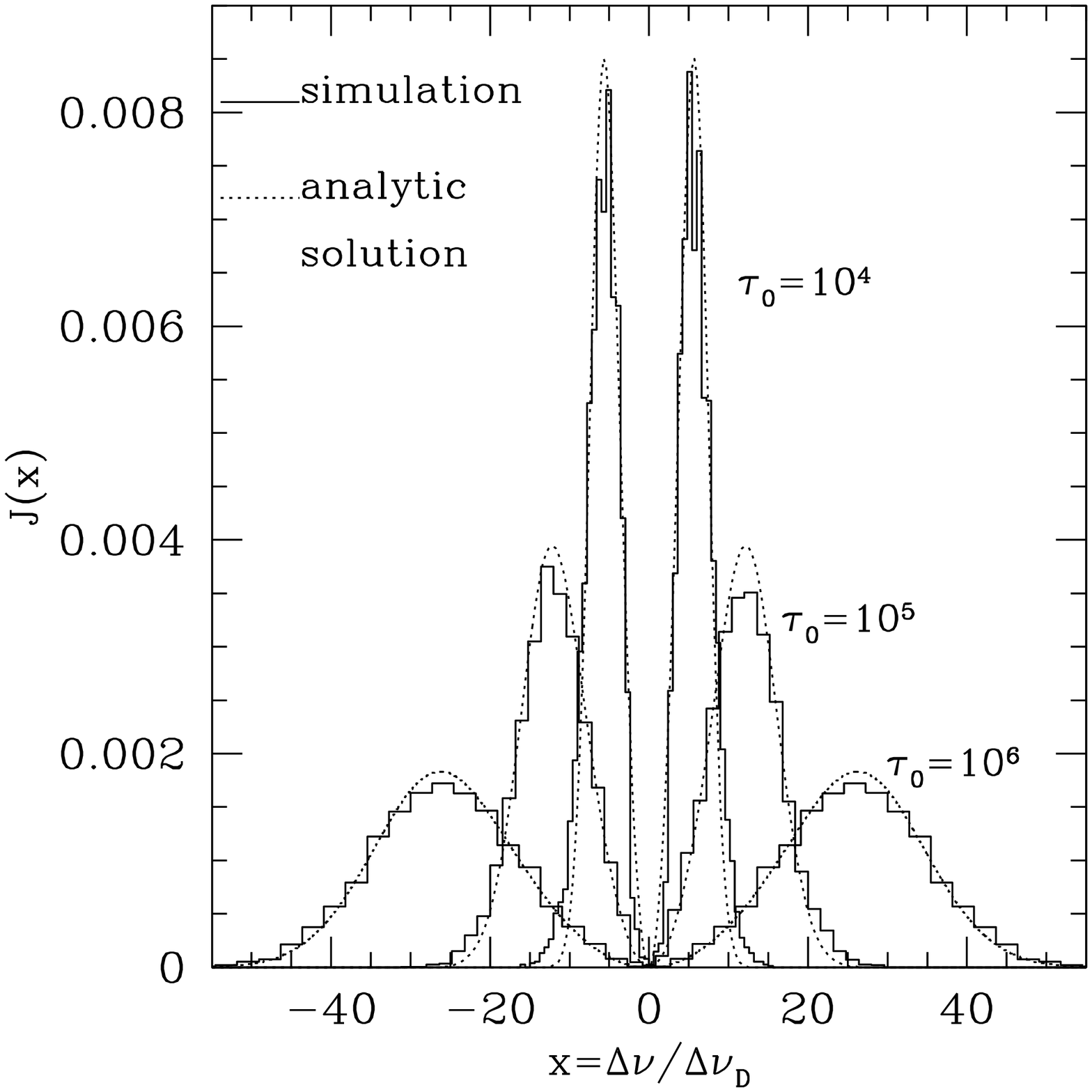}}\hspace{0.5cm}{\epsfxsize=3.5truein \epsffile{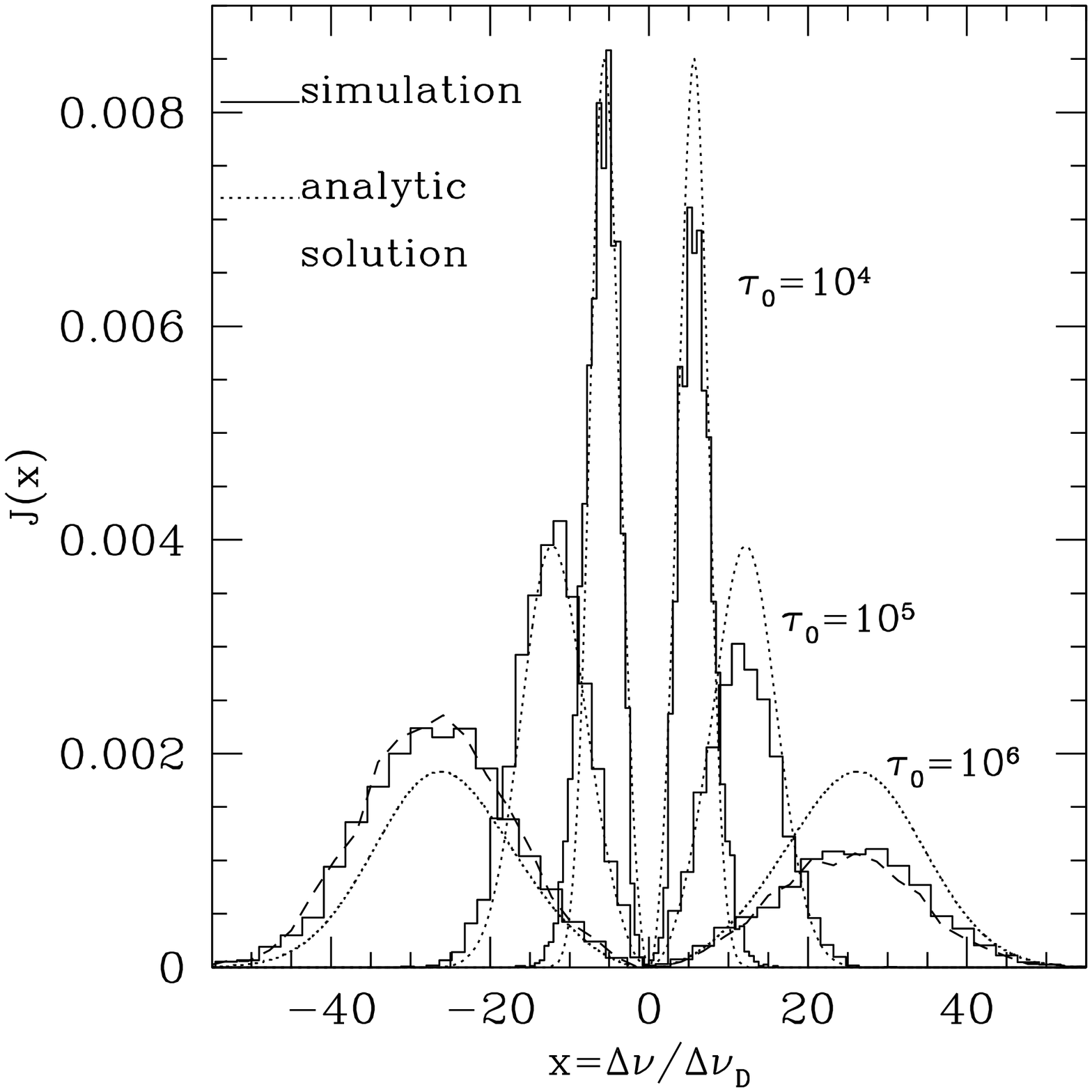}}} 
\caption[Comparison with Neufeld]{{\it Left panel:} Emergent spectra 
from the Monte Carlo RT ({\it solid histograms}) and as predicted analytically by \cite{neufeld90} 
({\it dotted lines}) for 3 different center--of--line optical depths. The agreement between Monte Carlo and analytical result
becomes better with increasing optical depth, as expected since the analytical solution is valid for very optically thick media.
{\it Right panel:} The same as in left panel but in this case the Monte Carlo results are derived with recoil being included,
whereas the analytic solution does not include recoil. The dashed line in the case of $\tau_{0}=10^6$ is obtained by modifying the
spectrum obtained from the Monte Carlo RT  without recoil by the factor correcting for recoil (see text for details).
\label{fig:neufeld}}
\end{figure*}

\subsubsection{\citet{neufeld90} test}
\label{sec:neufeld}
\cite{neufeld90}  derived an analytic solution in the limit of large optical depth  for a source radiating 
resonance line  photons in a  thick, plane-parallel, isothermal semi-infinite slab of uniform density. 
The analytic emergent spectrum as a function of frequency shift for a midplane source is
\begin{equation}
J(\pm \tau_{0}, x) = \frac{\sqrt{6}}{24} \frac{x^2}{\alpha \tau_{0}}\frac{1}{\cosh[(\pi^4/54)^{0.5} 
(|x^3-x_{i}^3|/\alpha \tau_{0})] }  
\label{neuf}
\end{equation}
with  $x=(\nu-\nu_{0})/\Delta \nu_{D}$, $\Delta \nu_{D}=\nu_{0} \sqrt{2k_{B}T/(m_{p} c^{2})}$ the thermal Doppler width, 
$x_{i}$   the injection frequency shift (zero for injection at line center), and $\alpha$ the ratio of the natural to two times the thermal
Doppler width.
The quantity $\tau_{0}$ is  the optical depth  from midplane to one
boundary of the slab at the line center.\footnote{Neufeld's definition, used in equation (\ref{neuf}),
is such that the optical depth at frequency shift $x$ is given as $\tau_{x}=\tau_{0} \phi(x)$.
Note that throughout this section, with the exception of equation (\ref{neuf}), 
 our  definition of $\tau_{0}$ is such that the optical depth at frequency shift $x$ is given
 as $\tau_{x}=\tau_{0} H(\alpha,x)$. 
 This definition was chosen following  recent studies \citep[e.g.][]{ahn_etal02b,zheng_miralda-escude02} so that comparisons
 with these studies be easier. Since
 $\phi(x)=H(\alpha,x)/\sqrt{\pi} $, our $\tau_{0}$ is smaller than
 Neufeld's  by a factor of $\sqrt{\pi}$. Note though that in the following sections  we return to the \citet{neufeld90}  definition of $\tau_{0}$.}  

This analytical solution is valid in the very optically thick limit, with the latter being defined according to \cite{neufeld90} 
 as $\tau_{0} \ge 10^3/(\sqrt{\pi} \alpha)$. This corresponds to  
 $\tau_{0}\ge 3.8 \times 10^4$ approximately for a temperature T=10 K  assumed in the tests we present here.
In deriving equation (\ref{neuf}) the scattering was assumed to be isotropic.
In addition, it was assumed coherence in the rest frame of the atom, an assumption that  makes the solution valid
at the low density limit only, as well as approximations under the assumption that wing scatterings dominate  were done (hence the solution is valid
at high optical depths). 
Furthermore, note that the classical slab  problem is independent of the real size of the slab (all quantities depend on $l/l_{0}$  with $l_{0}$ the actual size of the finite dimension). 
Lastly, for this solution it is assumed that the source has unit strength  and is isotropic, namely
it emits 1 photon per unit time or $1/ 4 \pi$ photons per unit time and steradian. 

For  center-of-line injection frequencies 
the emerging spectrum has maximum at $x \simeq \pm 0.88119 (\alpha \sqrt{\pi} \tau_{0})^{1/3}$, and an average 
number of scatterings $N\simeq 0.909316 \sqrt{\pi} \tau_{0}$ \citep[][with  $\tau_{0}$ 
in these expressions defined using our conventions rather than Neufeld's]{harrington73}.
This scaling of the mean number of scatterings with optical depth in the case of resonant-line RT
in extremely optically thick media  was first explained by \citet{adams72}, who understood that photons escape the medium after a series of excursions
to the wings. Before this study it was believed that 
the number of scatterings scales with $\tau_{0}^{2}$,  as would be predicted by plain  spatial random walk arguments \citep{osterbrock62}.  
We briefly review the  interpretation given by \citet{adams72} with respect to 
 the linear scaling of the mean number of scatterings with optical depth, since we  refer to it
extensively in the following sections.

The mean number of scatterings is  the inverse of the escape probability per scattering.
The escape probability per scattering is the integral of the probability per scattering that a photon is scattered 
beyond certain frequency shift $x_{*}$. \citet{adams72} identified this frequency as the frequency where the photon, while
performing an excursion to the wings, and before returning back to the core, travels an rms distance comparable to the size of
the medium. Note that this is in fact an essential difference in the understanding of resonant-line RT in extremely thick
media compared to the spatial random walk approach. The latter approach assumes that during an excursion to the wings the photon travels an rms distance
much smaller than the size of the medium. Thus, the first step is to determine $x_{*}$. 
Using the redistribution function (i.e., the function that gives the probability that a photon with certain frequency shift $x$ before scattering
will have a frequency shift $x^{'}$ after scattering)  
one can calculate both the rms frequency shift and the mean frequency shift  of a photon which is scattered repeatedly.
For a photon initially in the wings with a frequency shift $x$ \citet{osterbrock62} found that the rms shift is 1 and the mean frequency shift is
$-1/|x|$. For $x\gg1$, the mean shift is much smaller than the rms and the photon is undergoing a random walk in frequency with mean number of scatterings
 $\sim x^{2}$. In real space, the rms distance traveled is equal to the square 
 root of the mean number of scatterings times the mean free path. In the wings, the Voigt profile varies
 relatively slowly and the mean free path is $\sim 1/\phi(x) \sim x^{2}/ \alpha$ line center optical depths (we 
 only focus on the scalings here, hence constants of order unity 
 are dropped). 
Thus, the distance traveled is $x/\phi(x) \sim  x^{3}/\alpha$.  Setting this  rms distance equal 
to $\tau_{0}$ we get $x_{*} \sim (\alpha \tau_{0})^{1/3}$, which is in fact the scaling of the frequency shift where the
emergent spectrum takes its maximum value.  Thus, going 
back to the mean number of scatterings, the escape probability {\it per scattering} will be $\sim \int_{x_{*}}^{\infty} A(x) dx$
with $A(x)$ a function to be determined. According to the previous discussion 
  $x_{*}$ is the minimum frequency shift for which the photon during an excursion to the wings can travel an
rms distance at least equal to the size of the medium. If, for simplicity, one assumes complete redistribution the probability that a photon is found after
scattering with a shift between $x$ and $x+dx$ is $\phi(x) dx$. However, this is not the probability {\it per scattering}, since the photon will scatter
$\sim x^{2}$ before returning to the core. Thus, $A(x)$ is $\phi(x)/x^{2}$, and  $N_{sc} \sim \left[\int_{x_{*}}^{\infty} \phi(x)/x^{2} dx\right]^{-1}$,
with $\phi(x) \sim \alpha/x^{2}$ in the wings. Using the above expression for $x_{*}$ one obtains $N_{sc} \sim \tau_{0}$, with 
the constant of proportionality being of the order of
unity. 

The emerging spectra without and with recoil included are shown in the left and right panel of Figure \ref{fig:neufeld}, respectively.
A convergence test indicates that these results are robust if more than of order  $10^{3}$ photons are used. 
Referring to the left panel of the figure, the agreement between the results obtained with the code and
the analytic solution gets better at higher optical depths. 
As has been already mentioned, the analytic solution is derived after a series of approximations done 
on the assumption of optically thick media. For example,  when deriving the analytic solution 
the Voigt function is set equal to $\alpha /\pi x^2$.
Setting the Voigt function equal to this approximation in the code makes the agreement even better. 
The way the spectrum behaves for different $\alpha \tau_{0}$ is expected qualitatively: the higher the optical depth or the lower
the temperature (the higher the $\alpha$), the more difficult it is for the photons to exit the medium and  
the photon frequencies must move further away from resonance  to escape. Hence,
the peaks of the emerging spectrum occur at higher frequency shifts, and  the separation between the two
peaks becomes larger. The width of the peaks gets larger with larger $\alpha \tau_{0}$ in agreement with the dependence of the
optical depth on frequency (i.e., when in the less optically thick regime  core photons are relevant and the
optical depth goes as $e^{-x^{2}}$, whereas in the more optically thick regime  wing photons are more relevant, and there the
optical depth scales as $1/x^{2}$).

In the right panel of Figure \ref{fig:neufeld} we present numerical results when recoil is included, 
along with the analytical solution (as a guide) that does not include recoil.
As expected, including recoil shifts more photons to smaller (more red) frequencies. 
The magnitude of the effect can be understood as reflecting the thermalization of photons around frequency
$\nu_{0}$ \citep{wouthuysen52,field59}. This process modifies the photon abundance by $\exp(-x/x_{T})$, with $x_{T}=k_{B}T/h \Delta \nu_{D}$.  
Indeed, in the right panel of Figure \ref{fig:neufeld} the dashed line for $\tau_{0}=10^6$ is obtained by modifying by 
$\exp(-x/x_{T})$ the emerging spectrum obtained
from the simulation when no recoil is included.
These results are in agreement with the results and interpretations by \citet{zheng_miralda-escude02}. 

\subsubsection{\citet{loeb_rybicki99} test}
\citet{loeb_rybicki99} address the RT problem  in a spherically symmetric, uniform, radially expanding  
neutral hydrogen cloud surrounding  a central point source of
\lya photons.  No thermal motions are included ($T=0$ K). They  find that the mean intensity $\tilde{J} (\tilde{r}, \tilde{\nu})$ as a function
of distance from the source $\tilde{r}$ and frequency shift $\tilde{\nu}$  in the diffusion (high optical depth) limit is given by
\begin{equation}
\tilde{J}=\frac{1}{4 \pi} \left( \frac{9}{4 \pi \tilde{\nu}^3}\right)^{3/2} \exp\left(-\frac{9 \tilde{r}^2}{4 \tilde{\nu}^3} \right)
\end{equation}
with $\tilde{\nu}=\nu/\nu_{\star}$, $\nu=\nu_{0}-\nu_{photon}$, $\nu_{0}$ the \lya resonance frequency, and $\nu_{\star}$ the frequency  where the optical depth
becomes unity. The scaled radius, $\tilde{r}$ is equal to 
$r/r_{\star}$, with $r_{\star}$  the
physical distance where the frequency shift due to the Hubble-like  expansion  of the hydrogen cloud 
equals the frequency shift that corresponds to unit optical depth ($=\nu_{\star}$). 
\begin{figure}[htb]
\begin{center}
\includegraphics[width=9cm]{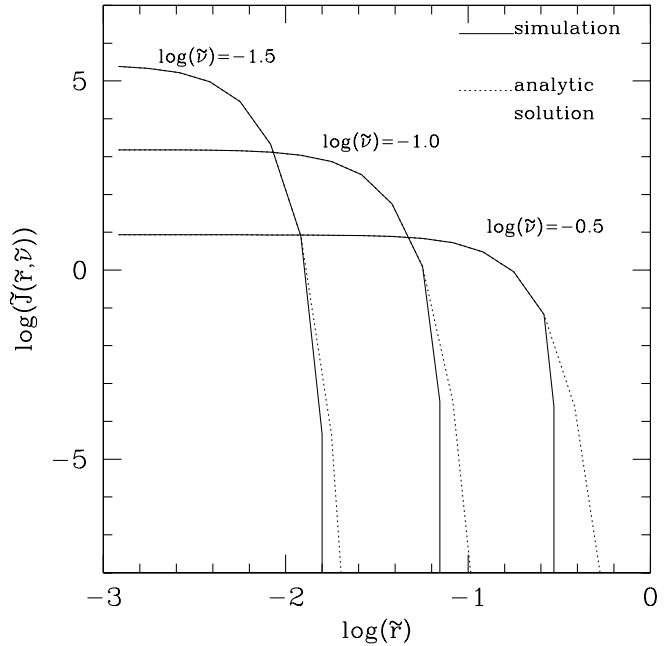}
\caption[Comparison with \citet{loeb_rybicki99}]{Mean intensity as a function of radius for certain frequency shifts. Solid lines show the results from the
Monte Carlo code and dotted lines show the analytic solution of \cite{loeb_rybicki99}, appropriate in the diffusion limit.
The specific frequency shifts plotted were chosen based on the fact that the diffusion limit is the right limit for $\tilde{\nu} << 1$
(for details and definitions  see  text).}
\label{loeb_rybicki}
\end{center}
\end{figure}
A comparison  of the results from the code with the analytic solution is shown in Figure \ref{loeb_rybicki}.
The analytic solution  becomes progressively more accurate the higher the optical depth (or the smaller the
frequency shift in the way this problem is parameterized, so that we are still in the core of the line). 
In addition, it  deviates  more and more from the (exact) simulation result at larger $\tilde{r}$, since the larger the $\tilde{r}$ the
more optically thin the medium  and thus the further away we are from the assumption of an optically thick medium made by  the analytic 
solution. Thus, the disagreement at high $\tilde{r}$ is real and not an artifact caused, e.g., by small number of photons
that would be inadequate to sample the low intensities at large $\tilde{r}$.

\subsubsection{Simple models of \lya emitters: Spherical clouds of uniform density and temperature}
\label{simple_models}
Here we  develop some simple models of \lya emitters. Even though there are no
analytic solutions for these cases, one could compare our results with the published results of \citet{zheng_miralda-escude02}.
More speficically, in this section, following these authors we model spherical neutral hydrogen clouds. We consider two different cases as far as the emission is concerned. In the first case it is assumed that we have a spherical cloud
with a \lya emitting point source at its center. In the second case we assume uniform emissivity, namely a photon is equally likely
to be emitted from any point within the cloud.
For each one
of these two cases we make runs assuming  the cloud is static, contracting and expanding. In the latter two cases the
contraction/expansion is assumed to be Hubble-like, namely the velocity  of the neutral hydrogen atoms scales linearly with the radius measured
from the center of the cloud. This velocity is set equal to 200 km/s at the edge of the system (and is negative/positive in the case of 
contraction/expansion).  For each case we perform two runs, one with column density equal to $2 \times 10^{18} \rm{cm}^{-2}$, typical for Lyman limit
systems, and one with column density equal to $2 \times 10^{20} \rm{cm}^{-2}$, typical of Damped \lya systems (or line center optical depths
equal to $8.3 \times 10^{4}$ and $8.3 \times 10^{6}$, respectively). 
In all cases the temperature is set equal to $2 \times 10^{4}$K. The initial photon frequency 
is assumed to be at the line center in the rest frame of the atom.  
In all results shown, 
the effect of recoil is included. Lastly, 1000 photons were used in all runs.

\begin{figure}[t]
\begin{center}
\includegraphics[width=7.5cm]{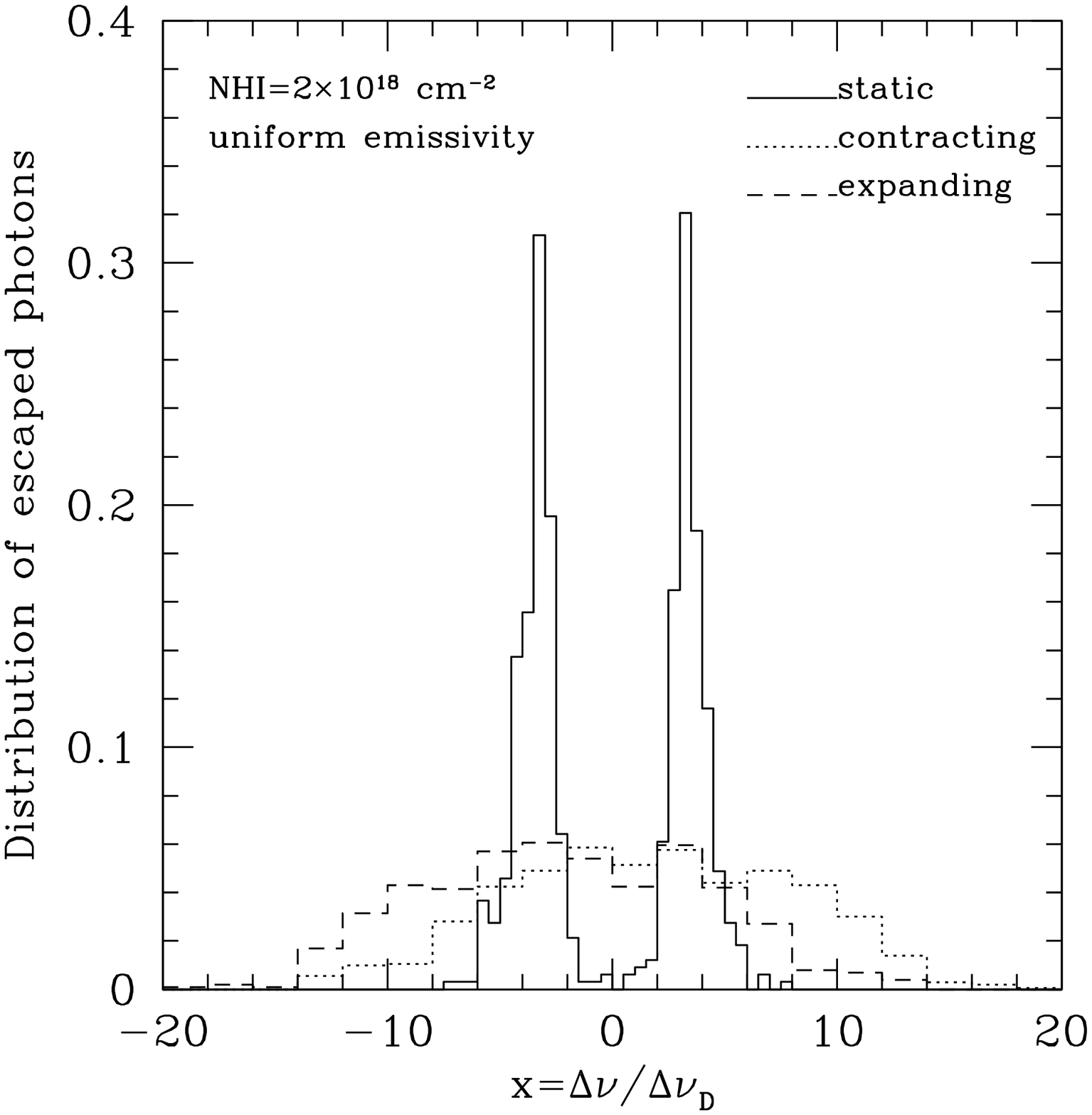}
\includegraphics[width=7.5cm]{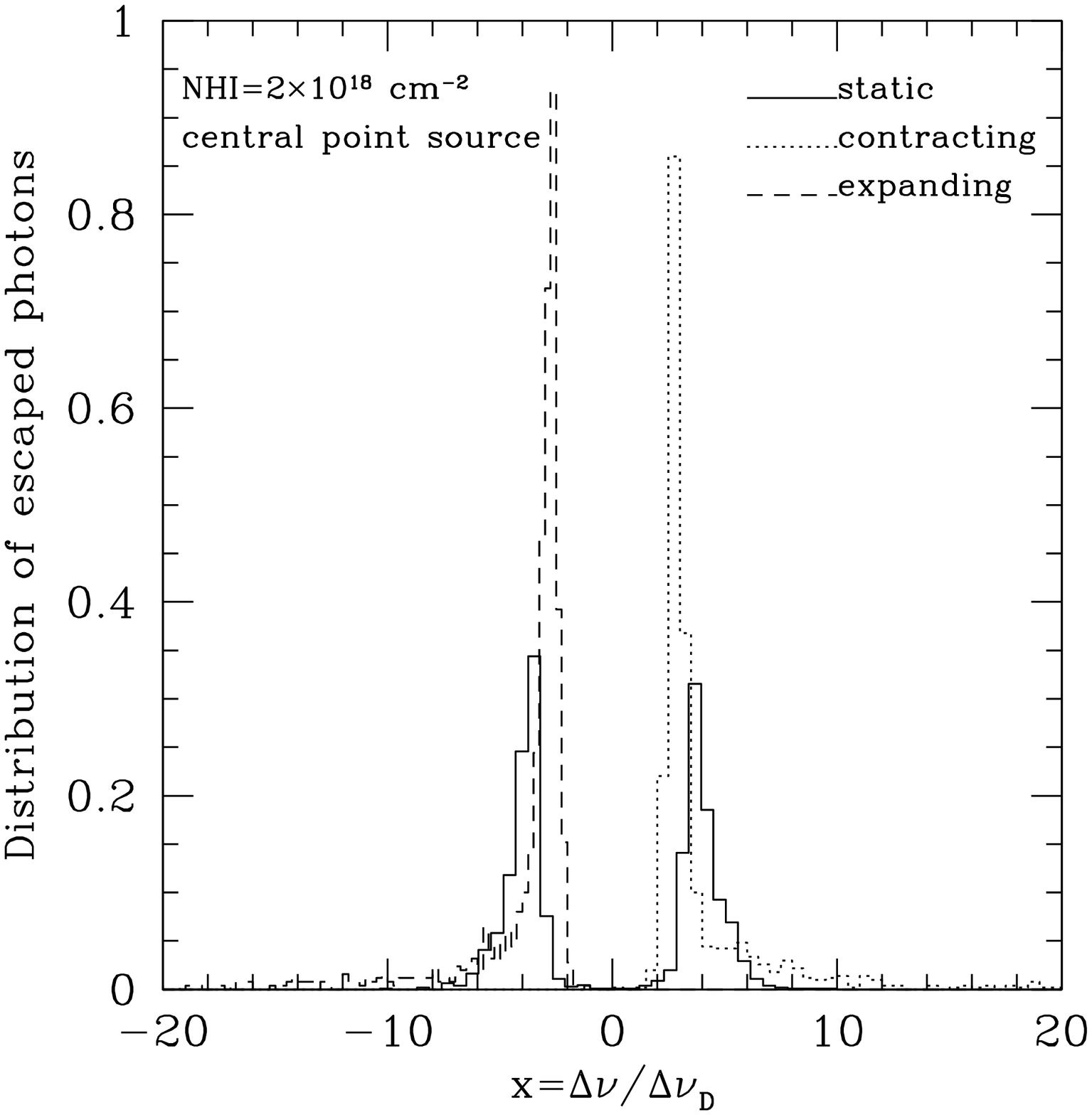}
\caption[Thin spherical cloud: uniform and central point source emissivity]{{\it Top panel}: frequency distribution of emergent \lya photons in the case of a static ({\it solid histograms }), a contracting 
({\it dotted histograms}), and an expanding ({\it dashed histograms}), isothermal, spherically symmetric neutral hydrogen cloud
with column density $\rm{N}_{\rm{HI}}=2 \times 10^{18} \rm{cm}^{-2}$ and  uniform emissivity. {\it Bottom panel}: same
as left panel but the \lya photons in this case originate from a central point source.}
\label{thin}
\end{center}
\end{figure}
\begin{figure}[htb]
\begin{center}
\includegraphics[width=7.5cm]{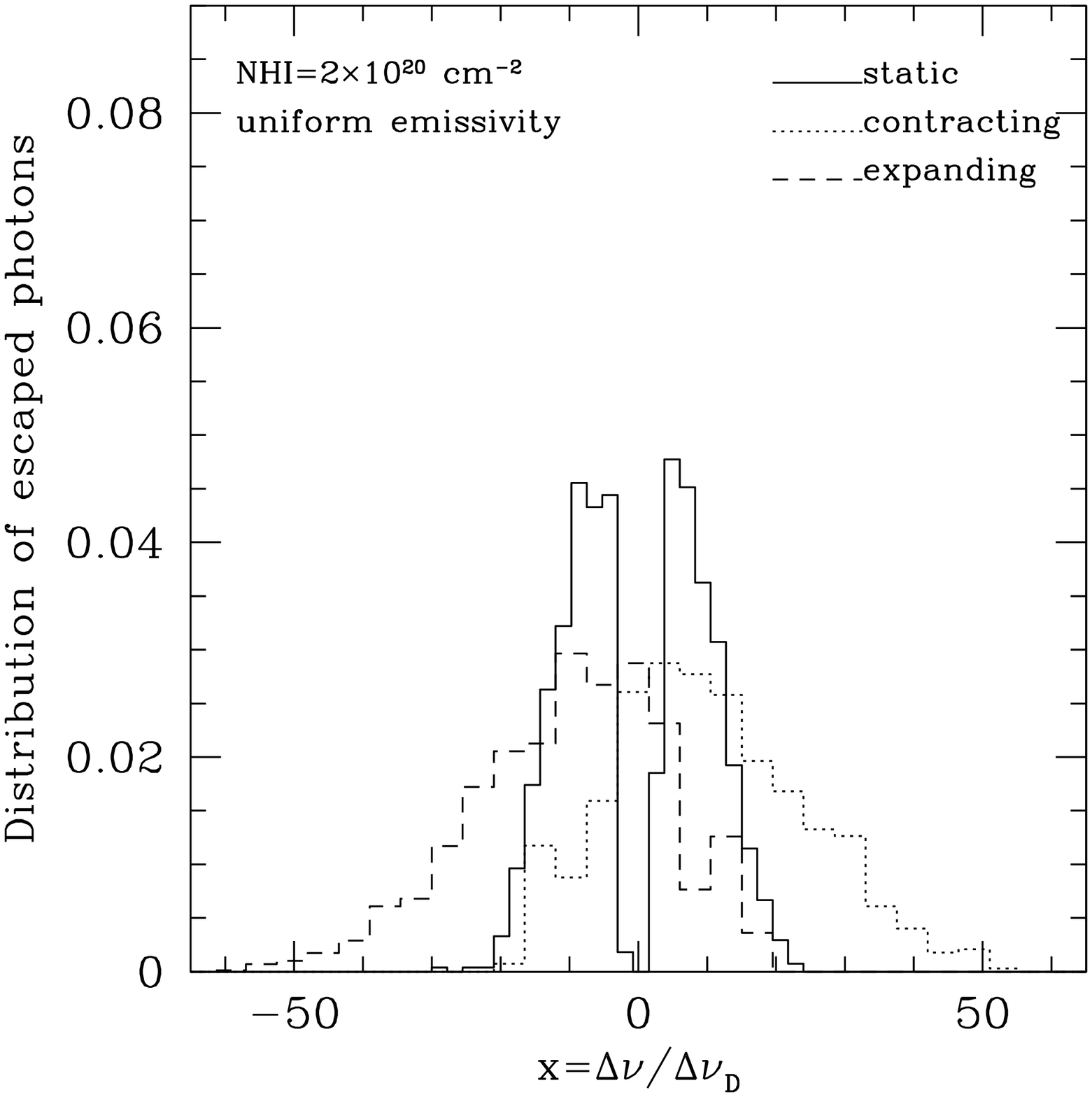}
\includegraphics[width=7.5cm]{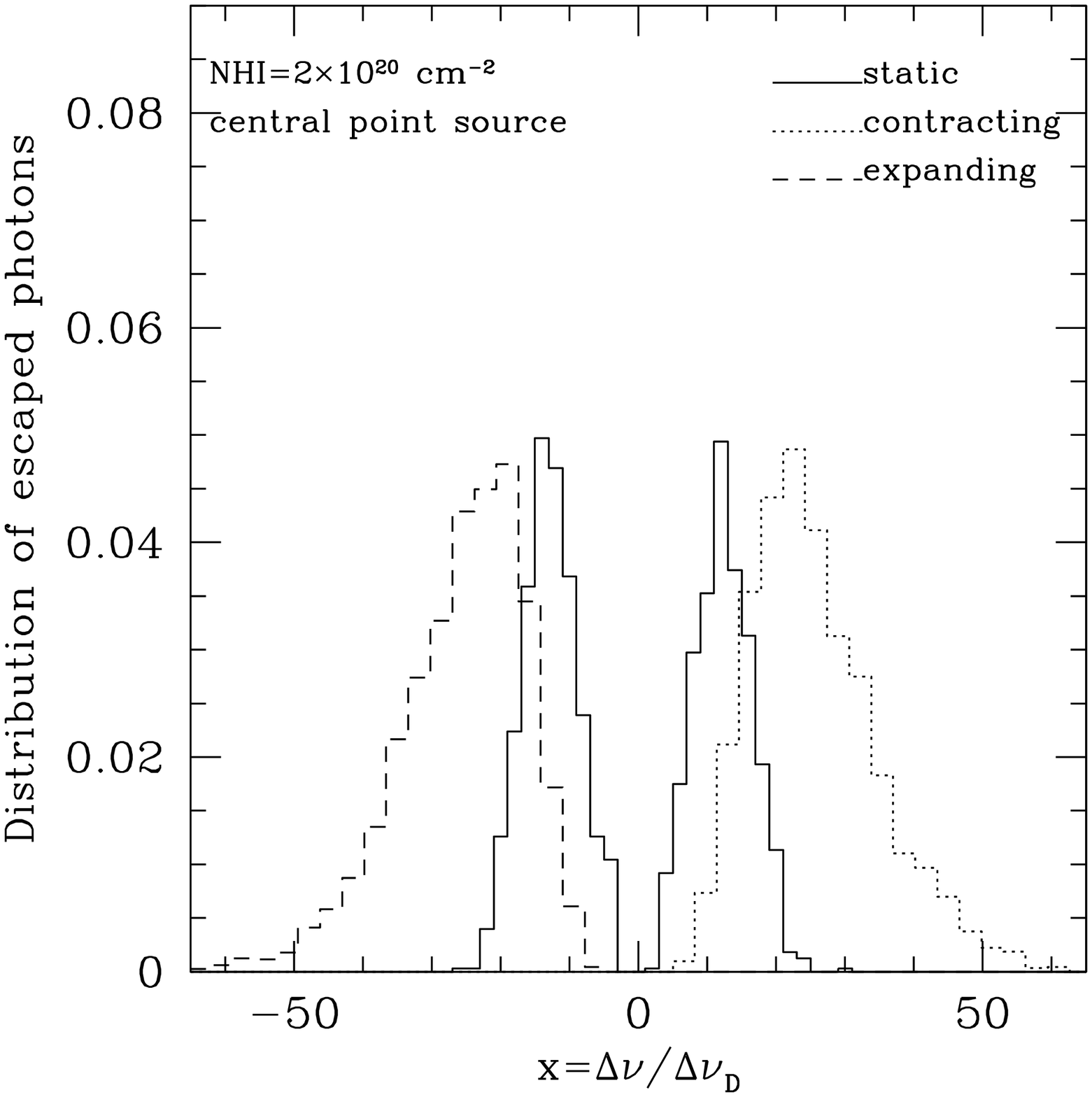}
\caption[Thick spherical cloud: uniform and central point emissivity]{{\it Top panel}: frequency distribution of emergent \lya photons in the case of a static ({\it solid histograms }), a contracting 
({\it dotted histograms}), and an expanding ({\it dashed histograms}), isothermal, spherically symmetric neutral hydrogen cloud
with column density $\rm{N}_{\rm{HI}}=2 \times 10^{20} \rm{cm}^{-2}$ and  uniform emissivity. {\it Bottom panel}: same
as left panel but the \lya photons in this case originate from a central point source.}
\label{thick}
\end{center}
\end{figure}
The results for the optically thin case  are shown in Figure \ref{thin} and that for the optically thick configuration 
are shown in Figure \ref{thick}. In both case the agreement with the results obtained by \citet{zheng_miralda-escude02} is very good. 
These  spectra can be understood qualitatively using the way  
the Neufeld solution  behaves depending on the optical thickness. 
In the case of an expanding cloud, the photons will escape on average with a redshift because they are doing work on the expansion of the
cloud as they are scattered.  Photons with negative frequency shifts (redshifted) can escape, but those with positive frequency shift (blueshifted) will be
scattered at some point and they have to undergo a series of many positive shift scatterings  to escape. Hence, the
blue part of the spectrum is suppressed. The situation is reversed in the case of a contracting cloud. It is important to keep in mind that the degree
of suppression of one of the two peaks due to bulk motions depends on factors such as the optical depth and the temperature.
In the case of uniform emissivity and expansion/contraction all spectra become broader because of the different velocities of the emission sites of the photons. In addition,
when the cloud is expanding (contracting), the blue (red) part of the spectrum is not suppressed as much as in the central point source case because, at least,
photons initially emitted close to the edge of the system have some chance of escaping even if they are blue (red).
In the optically thicker cloud, as soon as the photon reaches 
a sufficiently large $x$ it is not likely that it will be scattered by an atom with the right velocity to bring the photon into the line center.
Rather, the photon will get another random shift in frequency and will follow an excursion in frequency while at the same time it
diffuses spatially.  This along with the fact that the optical depth has a power law rather than Gaussian dependence on $x$ 
 broadens the peaks compared to those of the optically thin case, exactly as discussed for behavior of the Neufeld solution. 
 In addition, the emission peaks move further away from the
center compared to those from the optically thin case, since the photons have to be further away from the center of the line in order to escape
when the medium is optically thick.  

\subsection{Accelerating the RT}
\label{sec:accel}
The previous tests demonstrated that our basic Monte Carlo scheme works well for the simple test cases.
When using it in its simplest form in high resolution cosmological simulations, such as the ART simulations (see \S \ref{app_sims}), it takes unrealistic
running times in order to produce results with sufficient numbers of photons.
This is because in the case of  resonant RT in extremely optically thick media,  
a significant amount of time is spent on the relatively insignificant core scatterings. 
If we define the core through the frequency range where the Doppler profile dominates over the
Lorentzian wings, then roughly speaking the core is given by  $\alpha/ \pi x_{c}^{2}=e^{-x_{c}^{2}}/\sqrt{\pi}$, where
$x_{c}=(\nu_{c}-\nu_{0})/\Delta \nu_{D}$ and $\alpha$ is as defined previously. 
For a temperature of $10^{5}$ K, the core is roughly $x_{c} = 3.5$. Also, assuming complete 
redistribution,\footnote{ In
other words, assuming that the frequency distribution after scattering is independent of the frequency before scattering and 
is given by the line profile (i.e., the  source function is independent of frequency). The 
assumption of complete redistribution
was found to be pretty accurate for core photons \citep{unno52,jefferies_white60}. This is intuitively expected since, when in the core,
the photon frequency shift is small or comparable to the thermal velocities of the atoms. Thus, the latter can have
a significant impact on the frequency of the photon and in effect they redistribute it after each scattering according to the
line profile.} the probability per scattering for a core photon
to exit the core is $I/(I+\rm{erf}(x_{c}))$ with $I=2 \alpha/(\pi x_{c})$ and $\rm{erf}(x_{c})=\frac{2}{\sqrt{\pi}} \int_{0}^{x_{c}}e^{-t^{2}}dt$. 
That is, roughly, the photon will have to scatter $(I+\rm{erf}(x_{c}))/I=1+\rm{erf}(x_{c})/I \simeq 1+I^{-1} \simeq I^{-1}$ times before exiting the core. 
Using 
the above core definition, one finds that this is equal to $\sqrt{\pi} e^{x_{c}^{2}}/(2 x_{c})$ and keeping only the dominant dependence on $x_{c}$, this is roughly $e^{x_{c}^{2}}$ or
$\sim 10^{5}$  scatterings.
These scatterings are insignificant in the sense that they happen in such copious amounts, without being accompanied by significant 
spatial diffusion, since the latter
occurs mostly  through the wings. 

One way to advance photons in very high optical depths is to use the technique of the 
{\it prejudiced first scattering} \citep{cashwell_everett59}.
With this technique one biases the $\tau$ values toward larger values than the ones that would be drawn 
from  equation(\ref{taur}). More specifically,
$\tau$ is chosen to be uniformly distributed in $[0,\tau_{esc}]$, with $\tau_{esc}$ the optical depth for escape. Then one weights the photons by
$\tau_{esc}e^{-\tau}$ to correct for the fact that $\tau$ (i) is limited to be less than or equal to $\tau_{esc}$, and 
(ii) is assumed to be  uniformly distributed in the $<\tau_{esc}$ range. Using 
this technique however does not improve run time requirements to the extent we need and clearly more drastic acceleration methods are needed.  

Exiting the core  does not in general guarantee that the photons escape. 
In fact,  the photons 
may  return back to the core many times before escaping. This is not surprising since, as we will discuss, the maximum core frequencies that can be used
 are much smaller than $x_{*}$ discussed previously. Especially for extremely 
optically thick media ($\alpha \tau_{0} >10^{3}$), this in-- and out--of--the--core procedure is still
very expensive to follow.
Hence, we accelerate our RT scheme by implementing two different methods, depending on the center-of-line optical thickness of the cell  a photon
 finds itself in ($\tau_{0}$), as well as on the thickness of the cell for the specific frequency shift of the incident 
 photon ($=\tau_{0} \phi(x_{i})$ with $x_{i}$ the frequency shift of
 the incident photon). In fact we  parameterize the optical thickness of a cell not only via $\tau_{0}$, the line-of-center optical depth from the center of the
 cell to one of its edges, but rather via the product of $\alpha$ and $\tau_{0}$, motivated by the Neufeld solution. This parameterization 
turns out to be
 very good for media less optically thick than those the Neufeld solution applies to.
We discuss these two acceleration methods, as well as some additional acceleration techniques  in the following subsections.

\subsubsection{Extremely optically thick cells: Controlled Monte Carlo motivated by the  Neufeld solution}
\label{controlled}
This acceleration scheme is based on controlled Monte Carlo simulations of resonance RT 
in cells (cubes) with  several physical conditions, 
representative of the extremely optically thick cells in the simulations.
The idea is to  obtain  trends and best--fit functional forms for the
spectra emerging from thick cells. These  spectra can  then be used when running the code  so that 
instead of following  the scattering of the photons in detail, we can draw the frequency of the  photon emerging 
from a thick cell  using  the pre-calculated spectrum  appropriate for the physical conditions in this cell.
In principle, controlled Monte Carlo simulations can be used for any range of optical thicknesses.
We use it only in the extremely optically thick cells where
$(\alpha \tau_{0})_{eff}>2\times 10^{3}$ and  $(\tau_{0} \phi(x_{i}))_{eff}\gg1$ with $x_{i}$ the frequency shift of the incident 
photon.\footnote{We have used the index {\it eff} because, as is discussed later in this section having in mind an implementation of
the RT code for AMR simulations,  to decide
whether this method is applicable or not we create a mesh on top of the simulation mesh. In this new mesh, the photon is always at the center of a cell.
Then it is the 'effective' physical conditions in this new cell that are relevant when deciding if the acceleration method at hand is applicable or not. 
In the case of simulations without a cell structure, the index {\it eff} becomes redundant, since there is no initial mesh to begin with.}
We do that because we are motivating this method by
the Neufeld solution which is applicable only at the diffusion limit. 
The inherent cell structure of the AMR simulation outputs or the cell structure that can be generated for other cosmological codes, 
along with the resolution imposed isothermality and uniformity of each cell,
 are conducive to some kind of modification of the Neufeld solution.
In some sense, with the advent of cosmological simulations, the  
contemporary  analogue of the
extensively studied classical slab problem is the completely unexplored  
problem of resonance RT in a cube. This motivated a detailed study of the resonance RT 
problem in cubes where the reader is referred to for more 
details and results \citep{tasitsiomi05}. Here we only
summarize briefly some key results relevant to the current study.

As discussed in \S \ref{sec:neufeld}, the Neufeld solution was obtained under some assumptions.
To fit the controlled Monte Carlo spectra with a Neufeld type spectrum we have to investigate how sensitively 
the solution depends on these assumptions, as well as whether these  assumptions are valid in cosmological simulations. 
This is done in the following paragraphs.
\subsubsubsection{Choosing the exiting frequency}
\label{exit_frequency}
The exiting frequency of a photon entering an extremely optically  
thick cell is drawn by an emerging frequency distribution similar to the Neufeld solution (equation \ref{neuf}).
However, the Neufeld solution is derived for a semi-infinite
slab, whereas the simulation cells are finite cubes. Furthermore,  the  solution  assumes isotropic scattering, no recoil, which
anyway is negligible in the simulations, and does not include  velocities such as those associated with peculiar motions or the Hubble
expansion. Lastly, it  assumes that the source of the radiation lies within the slab,\footnote{ More
specifically, 
equation \ref{neuf}  assumes that the source is a plane source in the middle of the slab. Due to symmetry arguments, 
a plane source located at the middle of a slab is equivalent with respect to the spectrum of the emergent radiation 
to a central point source. \citet{neufeld90} provides a more general expression for different 
source positions.} and is valid for optically thick frequencies ($\tau_{0} \phi(x_{i}) \gg1$). 

\begin{figure}[t]
\begin{center}
\includegraphics[width=9cm]{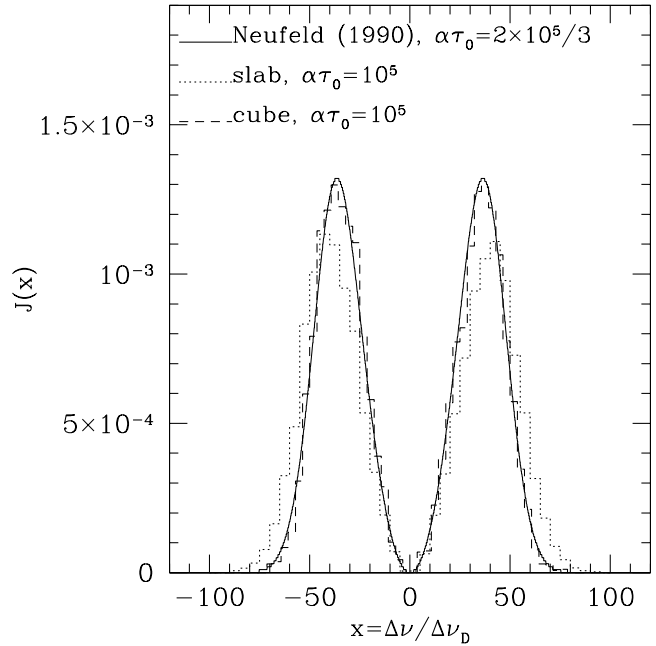}
\caption[Convergence wrt number of photons]{Comparison of the emergent spectra from a semi-infinite slab ({\it dotted histograms}) and 
a cube ({\it dashed histograms})
of the same physical conditions ($\alpha \tau_{0}=10^{5}$). Also shown is the analytical 
solution derived by \citet{neufeld90} ({\it solid line}) for the emergent spectrum from a semi-infinite slab.
Note that the analytical solution is for $\alpha \tau_{0}=2\times 10^{5}/3$, which is the 'effective'  $\alpha \tau_{0}$ 
 one has to use in the analytic solution obtained for a semi-infinite slab, for the solution to give the spectrum from a finite cube
 of the same physical conditions as the slab.
\label{slab_cube}}
\end{center}
\end{figure}

Starting from the point on bulk velocities, we use the Neufeld solution -- applicable  for an observer  moving with the
bulk flow of the fluid -- by taking into account the way the specific intensity
transforms between two inertial observers moving at a certain speed  with respect to each other (i.e., $I_{\nu}/\nu^{2}$ is invariant, where $I_{\nu}$ is 
the {\it number} of photons rather than the energy intensity. In the latter case,  the quantity that would be invariant would be $I_{\nu}/\nu^{3}$).   
The second point we address has to to do with the slab versus cube difference between the analytical solution and the simulations.
As discussed in \S \ref{sec:neufeld},  Neufeld's solution depends on one parameter,  $\alpha \tau_{0}$.
Qualitatively, one expects that  the spectrum emerging from a cube rather than a slab be  well described by the same
solution but for an effective $\alpha \tau_{0}$ smaller than the actual $\alpha \tau_{0}$ of the cell.  The 
reason for this
is that when for example observing the emergent flux from the z-direction in a cube, we lose all photons that in the case of the slab would wander, scatter
many times along the infinite dimensions and finally find their way out from the z-plane. In the case of the cube these photons would not be counted
simply because they have exited the cube from  planes other than the z-plane.  This would be equivalent to solving the problem that Neufeld solved but this
time including losses of photons (or, more appropriately, by generalizing the 2-dimensional diffusion equation derived by Neufeld into
a four dimensional one -- instead of $\tau, \nu$ now the  intensity will be a function of $\tau_{x}, \tau_{y},
\tau_{z}$ and $\nu$).
Numerical experimentation of RT in cubes and slabs of the same physical conditions, verified 
that the above guess is correct. In fact, the cube spectrum is well described by the Neufeld solution for a slab if $2/3$ of
the $\alpha \tau_{0}$ of the cube are used as input parameter to the slab analytic solution. 
This is shown in Figure \ref{slab_cube} \citep[also see][]{tasitsiomi05}.

Furthermore, the Neufeld solution assumes that the source of radiation lies within the slab (or cube in our case). In fact,
the version of the solution we have been discussing so far (equation \ref{neuf})  assumes that the source is at the center
of the slab.  
However, in the case of mesh--based codes, as photons cross from one cell to the other,
in general the source is  not at the center of the cell. For codes without an inherent cell structure the obvious solution to this is to create  a cell
and have the photon at each instant at the center of the cell. As  is discussed in what follows, this turns out to be the most efficient solution in the
case of mesh--based codes as well. 

Neufeld provides a more
general expression for various source positions within the slab, as well as for the transmission and reflection coefficients assuming an external source.
Using either option for mesh--based codes, trying to take advantage of the already existing mesh structure, creates complications: in the case of a non-central but internal source, the equivalence 
of a point or infinite plane source -- necessary for all the above discussion to be valid --  breaks down
if the source is not located at the center of the slab. And using the reflection/transmission probabilities makes the algorithm more complicated. 
But most importantly, there is an intrinsic limitation in the simulations due to finite resolution: it is not clear 
how meaningful it is to be discussing  differences
in position less than the cell size (i.e., if one can really tell the edge from the center of the cube). Instead, at every point the photon is found we 
create a new mesh on top of the simulation mesh. The photon is always found at the center of a cell whose physical parameters are calculated using the
cloud-in-cell weighting scheme. Each time the size of the cell is set to the simulation cell size the photon is in.  
Note that it is the physical parameters of this effective cell  that determine the way the code proceeds
(i.e., if the effective cell  $\alpha \tau_{0}$ is larger than $2 \times 10^{3}$ and $\tau_{0} \phi(x_{i}) \gg1$ 
then the controlled Monte Carlo results are used. If one of these two  conditions (or both) is not satisfied in
the effective cell  then the code returns to the original cell. Depending on the 
original cell  physical conditions and the photon frequency either the exact Monte Carlo or
the method described in \S \ref{skipping} is used).

In the Neufeld solution the condition  $\tau_{0} \phi(x_{i})\gg1$ allows him to truncate a  series appearing in the solution
process by keeping up to first order terms in $1/ (\tau_{0} \phi(x_{i}))$. Thus, the solution is valid only for optically thick injection frequencies.
We find that the higher order corrections are pretty small. However, for a certain tolerance, one must decide  how thick is thick enough for the Neufeld spectrum to be applicable.
We take that  the spectrum from a slab is satisfactorily predicted by the analytical solution
for frequency shifts for which $\tau_{0} \phi(x_{i}) \ge 10$. 

As has been shown in \S \ref{sec:neufeld} the recoil effect can be easily accounted for multiplying the Neufeld solution by the appropriate factor. 
In any event, the recoil effect for our conditions is negligible and hence is  dropped in the simulation calculations.
To see this, the recoil effect corresponds to a frequency shift that would be caused by a velocity $\simeq h \nu/m_{p} c
=$ 3 m/s. This velocity is negligible compared to the thermal velocities 
expected in cosmological simulations, and  given the peculiar and Hubble flow velocities, the small non-coherence in the atom's rest frame
introduced by recoil will be totally unobservable.
Hence, the Neufeld approximation is good in that respect as well.

\subsubsubsection{Choosing the exiting direction and point} 
Referring to $\mu$, the cosine of the angle with which the photon is exiting a cell, measured with respect to the normal to the exiting surface,
we draw its value from the following cumulative probability distribution function (cpdf) \citep{tasitsiomi05} 
\begin{equation}
P(<\mu)=\frac{\mu^{2}}{7}(3+4\mu) \, .
\label{dir}
\end{equation}
This cpdf is found to be an excellent description of the directionality of the emergent spectrum and 
clearly deviates from isotropy.
In fact, it verifies the findings of other studies that in optically thick media photons tend to exit in directions perpendicular to 
the exiting surface \cite[see, e.g.,][]{chandrasekhar,phillips_meszaros86,ahn_etal02b}. In the case of RT in accretion disks this has been identified
as an expected limb darkening \citep[or 'beaming';][]{phillips_meszaros86} of the disk (i.e., the disk is very bright when observed face on and less bright when observed edge on).
In cases of very optically thick media, the emerging radiation directionality approaches the Thomson
scattered radiation emergent from a Thomson-thick electron medium. This Thomson limit obtained initially by \citet{chandrasekhar}, was confirmed later
numerically by \citet{phillips_meszaros86}. 

It has been implied by some authors \citep{ahn_etal02b} that 
the fact that in optically thick media RT occurs mostly  via wing photons with
the latter being described by a dipole phase function (see \S \ref{scattering}),  and the fact that Thomson scattering  is also described by a Rayleigh
(dipole) scattering phase function, explains why the resulting $\mu$ probability distributions are similar. 
However, we find the same cpdf 
when the scattering is taken to  follow either an isotropic or a dipole distribution. 
For such optical thicknesses the details of the exact phase function do not matter, at least not
with respect to the exiting angle cpdf. All the  phase functions involved in \lya  scattering are only mildly anisotropic
and they simply  enhance a  little bit the coherence of the scattering at the observer's frame compared to the isotropic
scattering case. So the fact that the exiting angle cpdf in extremely optically thick slabs (cubes) 
does not depend crucially on the assumptions on the phase functions does not
come as a surprise.
The underlying physics is simply that   in extremely thick media 
most  of the photons escape along the normal to the slab where the opacity is smaller.
The azimuthal angle $\phi$ with which the photon exits a cell is distributed fairly uniformly in $[0,2\pi]$ \citep[for more details see][]{tasitsiomi05}.

Referring to the distribution of exit points, one can argue that trying to specify the exact coordinates of the exit point of a photon 
from a simulation
cell is, in some sense, superfluous since there is always the resolution  limitation.  
Thus, we assume that the 
exiting points are distributed uniformly. 
The deviations of the exiting points from uniformity are relatively  small \citep{tasitsiomi05}.
Similarly, resolution limitations  make us focus on total distribution functions of photon properties -- where total here means distributions averaged over an entire cube side -- without 
regards to a possible dependence of these distribution functions on the photon exit point.    

Lastly, we have checked whether the emergent photon parameters can be drawn independently. We found  no significant correlations among them
(e.g., we checked for correlations between emergent frequency shift and  (preferred) range of exiting directions). Thus,  drawing 
them independently
is correct. 

\subsubsection{Moderately optically thick cells: Skipping the  core scatterings}
\label{skipping}
This acceleration scheme is used if the cell the photon is in has 
$1 \leq \alpha \tau_{0} \leq 2 \times 10^{3}$. It is also used in the case of cosmological simulation
codes with a pre-existing mesh when the cell the photon is in 
has $\alpha \tau_{0}> 2 \times 10^{3}$, but the effective cell (see \S \ref{controlled}) has
$1 \leq \alpha \tau_{0} \leq 2 \times 10^{3}$, and thus the previous acceleration scheme (discussed in \S \ref{controlled})
is not applicable.
The scheme is based on the idea  that if a photon is within a certain {\em core} (to be determined), we 
can skip all the core scatterings and go directly to the scattering with 
a rapidly moving atom that can bring the photon out of the core \citep[for some first implementations of this idea see][]{avery_house68,ahn_etal02b}. As 
soon as this happens, the 
initial detailed transfer resumes until either the photon escapes or re-enters the core. 
The scheme's validity relies  upon the correct choice of the core value, so that 
while in the core the photon does not diffuse significantly
in space, whereas significant diffusion occurs when the photon exits the core.

To achieve the scattering that brings the photon outside the core 
 we choose thermal velocities (in units of $\sqrt{2kT/m}$) from the distribution \citep{avery_house68,ahn_etal02b}
\begin{equation}
p(v)=\frac{1}{\sqrt{\pi}} e^{-v^{2}}
\label{eq:core}
\end{equation}
and in the range $[v_{min},v_{max}]$. The lower limit  $v_{min}$ is the minimum velocity necessary 
for the photon to just make it 
to the core $x_{c}$. The upper limit is formally infinite,  but for any practical realization it can be set to a large enough number (e.g., 
$\sqrt{x_{c}^{2}+10}$). 
For a scattering to  bring the photon to just $x_{c}$ from the center, 
independent of the directions of incident and outgoing photon, and under the assumptions of 
 coherence in the rest frame of the atom,  isotropic scattering phase function, and  
 zero radiation damping, it can be shown that $v_{min}=max(|x|,|x_{c}|)$ \citep{hummer62}, with
$x$ the initial frequency shift (as usual in units of the thermal Doppler width). In our case 
it is always $v_{min}=|x_{c}|$ since the
photon is inside  the core.  We checked and verified that
the assumptions under which  $v_{min}$  is derived  
are good for cosmological  simulations.
This is not surprising since, e.g., the assumption  of 
an isotropic phase function is not very crucial. As discussed already, 
none of the relevant phase
functions is strongly anisotropic. Those that are anisotropic simply tend to favor slightly smaller  frequency shifts (since they favor
post-scattering directions close to pre-scattering directions) and hence increase a little bit the
coherence in the observer's frame from scattering to scattering. At the limit of many scatterings (and while still at the 
optically thick regime) this is not a significant effect \citep[for the tiny differences in the frequency redistribution function with isotropic versus
dipole phase function see Figure I of][]{hummer62}.
Or, the assumption of coherence in the rest frame of the atom is also expected
to be a pretty good assumption for the 
media in the simulations from the point of view of the recoil effect, as we discuss in \S \ref{exit_frequency}, and from the point of view of
collisions as we discuss in \S \ref{sec:collisions}.

To motivate the core values we can use (i.e., the maximum frequency shifts
for which we can ignore the repeated scatterings without biasing the results) we must take into
account the different physics
of  resonant RT in the two different regimes, $1\le \alpha \tau_{0}\le 2\times 10^{3}$ and
$\alpha \tau_{0} > 2 \times 10^{3}$. In the first regime photons escape on a {\em single longest flight} \citep{adams72} in accordance with the understanding of resonant RT in moderately thick media
developed by \citet{osterbrock62}. In this thickness regime the important frequency is the frequency where the optical depth becomes unity. Photons within this frequency shift barely diffuse in space, whereas as soon as they exit this frequency shift they escape while taking their longest spatial step ({\em flight}).
In the second, extremely optically thick regime ($\alpha \tau_{0}>2 \times 10^{3}$)  as \citet{adams72} suggested, photons escape during a {\em single longest excursion} rather than flight. In this case the important frequency is the frequency with the following property: if a photon is given  this frequency and is left to slowly return to the center of the line
(by performing a double random walk, in space and frequency),  the overall rms distance that it will 
travel in real space while returning to the line center equals the size of the medium (i.e.,  the important frequency shift in this case is the shift $x_{*}$  discussed in  \S \ref{sec:neufeld}).  This physics motivates our cores, i.e., for moderately thick media the core must be safely optically thick, whereas for extremely optically thick media the core must be safely smaller than $x_{*}$. Then using numerical experimentation we find the exact maximum possible core values that can be used 
in each case.

\begin{figure}[htb]
\begin{center}
\includegraphics[width=7.5cm]{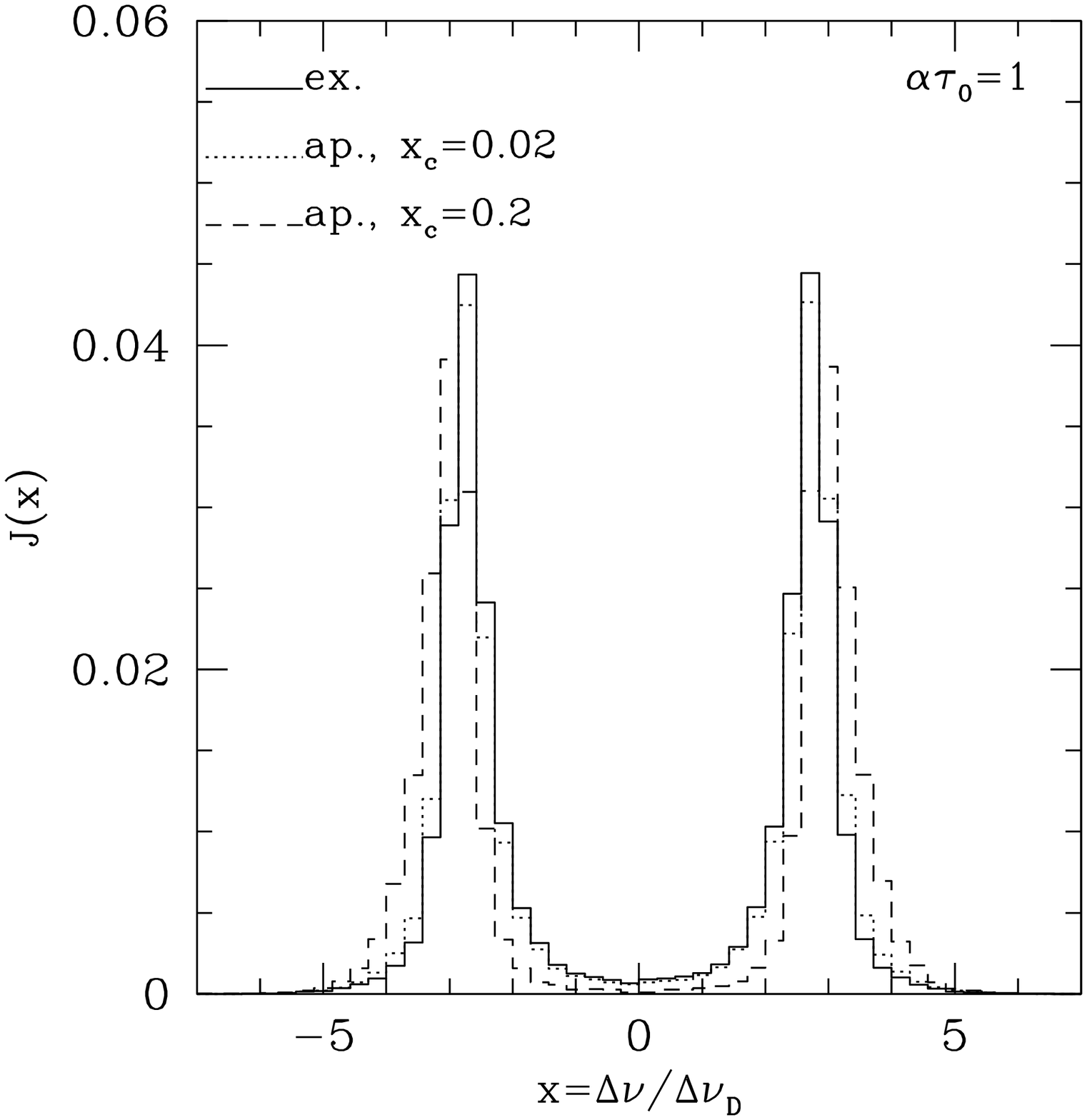}
\includegraphics[width=7.5cm]{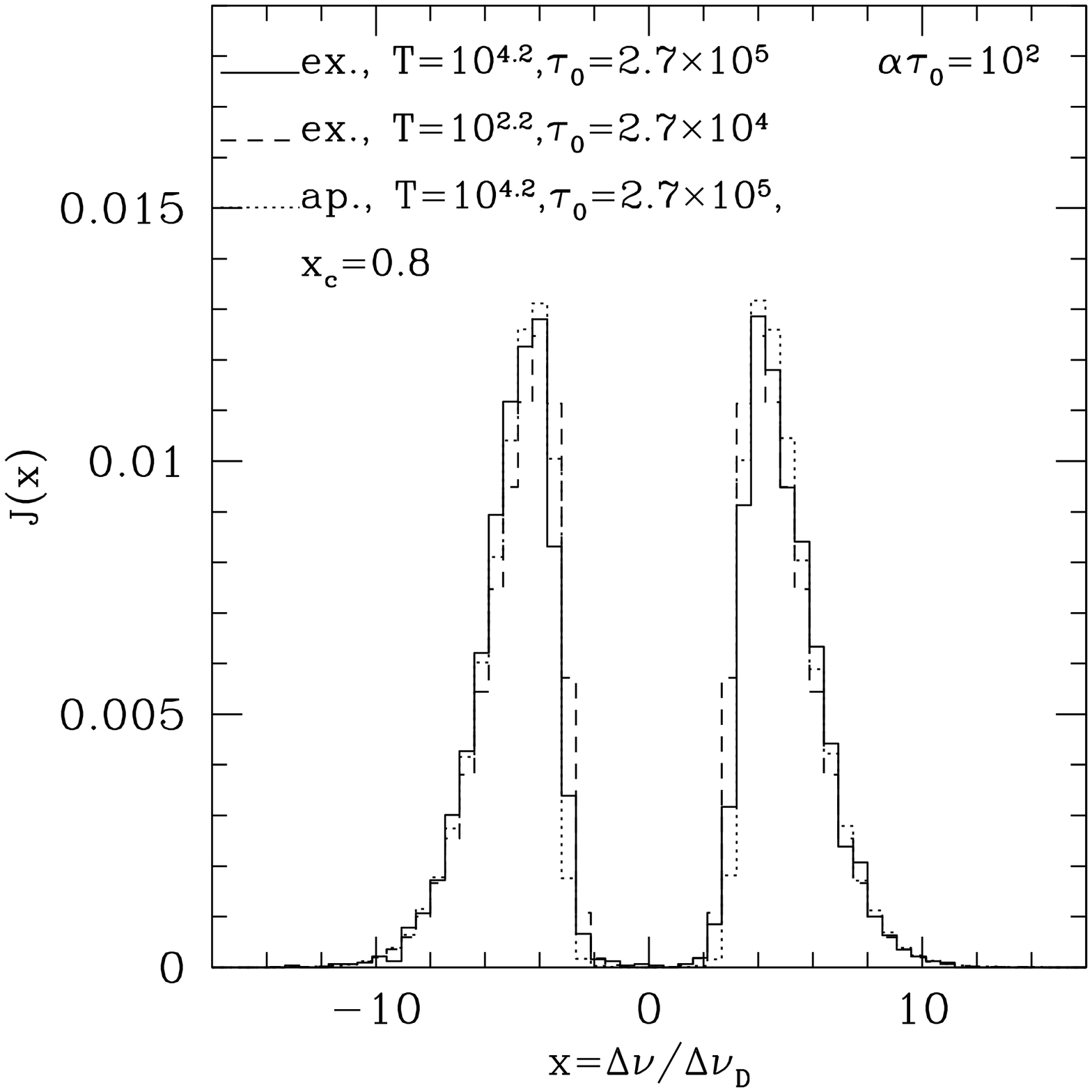}
\caption[]{{\it Top panel:} Comparison of the exact Monte Carlo results ('ex.', {\it solid line}) and the results obtained using the 
core
acceleration method ('ap.', {\it dotted line}) for the minimum cell $\alpha \tau=1$ for which this acceleration method is used in the \lya RT code.
Also shown is a larger core frequency, $x_{c}=0.2$, which shows the way the emergent spectrum is biased  if one uses a higher core frequency. 
{\it Bottom panel:}  Same as in top panel but for more optically thick cells, $\alpha \tau_{0}=10^{2}$. In this case a core frequency $x_{c}=0.8$ can be used.
In addition, we  show exact results for a different pair of temperature and optical depth ({\it dashed line}), that however correspond to $\alpha \tau_{0}=10^{2}$.
Clearly, $\alpha \tau_{0}$ is a good way to parameterize the problem at these moderate optical thicknesses.
\label{core}}
\end{center}
\end{figure}

A comparison of the exact Monte Carlo and the core acceleration scheme applied to moderately thick media is shown 
in Figure \ref{core}. Note that these spectra are  {\it one} cell runs, and are not the final results of the RT around the \lya emitter (which 
are discussed in a later section).  
In  the top panel, we present the exact emergent spectrum from a cube with
$\alpha \tau_{0}=1$, as well as the spectrum obtained if a core $x_{c}=0.02$ is used. Despite it being 
a pretty small core, it improves the speed of the algorithm
by orders of magnitude.\footnote{The exact improvement factor depends on optical thickness, and is higher for thinner cells. Furthermore, the improvement
factor is different for the same $\alpha \tau_{0}$ but different temperatures and optical depths. More specifically, it is
 higher for lower optical depths and
temperatures.}  Also 
shown is what the bias would be if one
used a higher core frequency ($x_{c}=0.2$): photons would be artificially shifted at higher (absolute)  frequency shifts. 
To find the maximum core that can be used
without this biasing, we made runs with successively higher cores. We use as cores: 0.02 for $1\leq \alpha \tau_{0}<10$, 0.1 for
$10\leq \alpha \tau_{0}<10^{2}$ and 0.8 for $10^{2}\leq \alpha \tau_{0}<2\times 10^{3}$. 
One can easily verify that for a wide temperature range these cores are safely within the optically thick regime.

The comparison between the exact Monte Carlo and  the accelerated
scheme for optically thicker cells (but still at the moderately thick regime) is shown at the bottom panel of Figure \ref{core}. 
We have seen via the Neufeld solution  
that characterizing a slab -- or a cube in our case -- using $\alpha \tau_{0}$
is very good in the case of very optically thick media ($\alpha \tau_{0} \geq 10^{3}$). In the bottom panel of Figure \ref{core} 
we present two different sets of temperature and $\tau_{0}$, which nevertheless correspond to the same $\alpha \tau_{0}$ (and smaller than
that for which the Neufeld solution is applicable). Clearly, $\alpha \tau_{0}$ 
parameterizes nicely enough these emergent spectra as well. This fact justifies our classification of simulation cells  with respect to their
$\alpha \tau_{0}$ value.  Note that the fact that the emergent spectrum for these physical conditions seems to depend on $\alpha \tau_{0}$ 
 is not trivial, and was checked only for ranges of temperature and optical depth that are 
anticipated to be relevant to cosmological simulation environments. A simple way to see why this may not be a general statement comes from the physics of RT in moderately thick media.  
As discussed in such media photons escape roughly when they reach the frequency where the optical depth is unity. 
If, for example, the frequency shift $x$ where the optical depth becomes unity is within the Doppler core (as anticipated) then this frequency shift is 
defined through $\tau_{0} e^{-x^{2}}=1$ and clearly depends only on $\tau_{0}$ and not on temperature. This is in contrast to extremely optically thick media where the 
frequency shift relevant for escape through the single longest excursion is $x_{*}\sim (\alpha \tau_{0})^{1/3}$ (see \S \ref{sec:neufeld}), namely it depends on $\alpha \tau_{0}$.

\begin{figure}[htb]
\begin{center}
\includegraphics[width=9cm]{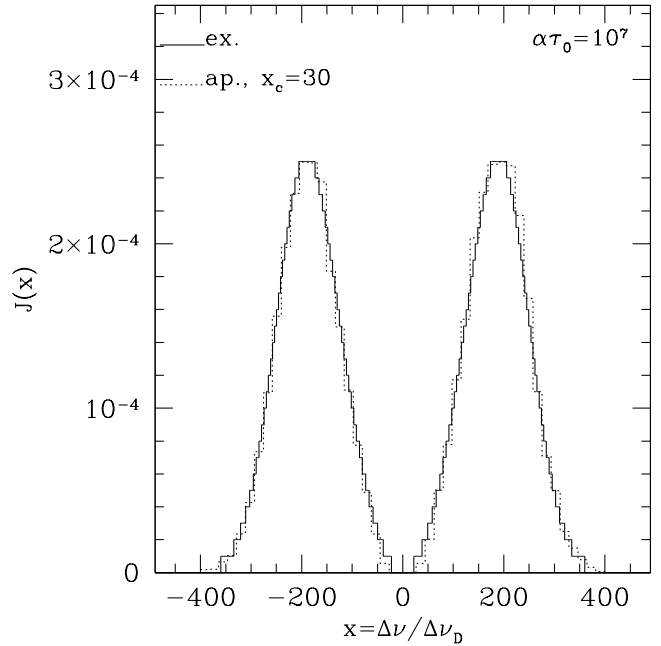}
\caption[]{Comparison of the analytic solution obtained by \citet[]['ex.', {\it solid line}]{neufeld90}  and the results obtained using the 
approximate core
acceleration method ('ap.', {\it dotted line}) for $\alpha \tau=10^{7}$ and $x_{c}=30$. The point of the figure is that at extremely high optical depths
the core values one can use can be pretty high.
\label{deep}}
\end{center}
\end{figure}

 
 In the case of extremely thick media we find roughly the following  maximum possible cores:
3 for $2 \times 10^{3} \leq \alpha \tau_{0}<10^{4}$, 5 for $10^{4}\leq \alpha \tau_{0}<10^{5}$, 7 for $10^{5} \leq \alpha \tau_{0}<10^{6}$, 
17 for $10^{6} \leq \alpha \tau_{0}<10^{7}$, 30 for $10^{7} \leq \alpha \tau_{0}< 10^{8}$,  and 80 for $\alpha \tau_{0} \geq 10^{8}$. 
As an example, in Figure \ref{deep} we show the Neufeld prediction for the emergent spectrum from a slab with
$\alpha \tau_{0}=10^{7}$ and the results of our acceleration scheme using a core $x_{c}=30$. This is a quite large core frequency, and still the
acceleration scheme gives a very accurate emergent spectrum.
The core values we find  scale with $x_{*}$ roughly as $x_{c} \simeq 0.15 x_{*}$. 

The  $\alpha \tau_{0}$-dependent core frequencies that we
 motivate here based on the different  
physics for different $\alpha \tau_{0}$ regimes is a quite new approach. Previous studies \citep[e.g.,][]{hansen_oh05}  define the core frequency
 as
the frequency where the wings start dominating over the Doppler core. 
Clearly, to achieve 
the best efficiency of the  
acceleration scheme, which is highly desirable in our applications due to the very complex environments, we have to use
a depth--dependent core definition. 
Other authors who considered variation of the core frequency with temperature and optical depth \citep{ahn_etal02b} 
find a bit different values than ours, at least for the low $\alpha \tau_{0}$ range that they worked with: they find that
a core frequency of about $\sqrt{\pi}$ can be used for $\alpha \tau_{0}>10^{3}$, with slightly higher values permitted for even larger 
$\tau_{0}$. However, we find that this value is a bit large for $\alpha \tau_{0} \simeq 10^{3}$, and that significantly higher core values can be used for
higher $\tau_{0}$. The reasons for our disagreement with \cite{ahn_etal02b} are not clear.  

The discussion with respect to the validity of this acceleration scheme has been limited so far to the emergent spectrum of radiation.
One would expect to see that indeed the assumption that the photons do not move significantly in space during the multiple core scatterings that are skipped
is true. And, that all other quantities, such as exit point and exit angle distributions remain the same, in addition to the emergent spectrum. 
The latter has been tested and found true. Furthermore, note that the angle information is relevant mostly when the photon is at the optically thin regime,
where anyway we use the exact transfer scheme. With respect to the exit points, or distances that the photons move while in the core, since these are not
larger than one cell size, limitations due to the
finite simulation resolution render these concerns moot.  To get an idea, following an argument similar to that presented in \S \ref{sec:neufeld} leading to $x_{*}\simeq (\alpha \tau_{0})^{1/3}$, and using 
the scaling $x_{c} \simeq 0.15 x_{*}$ one finds that by ignoring the scatterings within the core for extremely optically thick cells roughly one ignores a spatial diffusion 
of the photons of order $10^{-3}$ of the size of a simulation cell. 
 
In summary,  each time a photon enters a simulation cell, there are the following three possibilities:
\begin{enumerate}
\item If the cell has $\alpha \tau_{0}<1$, the exact Monte Carlo RT is used. 
\item If the cell has $1 \leq \alpha \tau_{0} \leq 2 \times 10^{3}$ and the photon frequency shift is $|x| \leq x_{c}$, then we 
skip  the core scatterings. If the photon frequency is outside the core we use again the
exact Monte Carlo RT.
\item If the cell has  $\alpha \tau_{0} > 2\times 10^{3}$ then if there is no pre-existing mesh structure of the
cosmological simulation then if the frequency of the photon is such that $\tau_{0} \phi(x) \gg 1$
then the controlled Monte Carlo results are used.
If there is a pre-existing mesh structure, then if $\alpha \tau_{0} > 2\times 10^{3}$
then the physical conditions of the effective cell are calculated.
If for the effective cell it is (i) $\alpha \tau_{0} > 2 \times 10^{3}$  and the frequency of the photon is such that (ii) $\tau_{0} \phi(x) \gg 1$, then
we use the controlled Monte Carlo motivated by the Neufeld solution.   
If  either (i) or (ii) is not true, then the first acceleration scheme is tried for the {\it original} rather than the {\it effective} cell.
And if it is not applicable, then the exact Monte Carlo scheme is used.
\end{enumerate}

\subsubsection{Calculating images and spectra} 
To construct images of the \lya emitters for various directions of observation the code
calculates the contribution to the image along a certain direction at {\it each} scattering \cite[see, e.g.][]{yusef-zadehetal84,zheng_miralda-escude02}. This contribution 
is $e^{-\tau_{esc}} P(\phi,\mu)$ where
$\tau_{esc}$ is the optical depth for escape from the current scattering position 
along the direction of observation to the observer, $\mu$ is the cosine of the angle between the direction of the
incident photon and the direction of observation, $\phi$ is the azimuthal angle,  and $P(\phi, \mu)$ is the normalized 
probability distribution for the photon direction (in fact
$P$ is independent of $\phi$ in our case). 

This way of calculating images and spectra 
has the advantage of giving  fairly good statistics for relatively small numbers of photons. Thus, by
lowering the number of photons needed for the results to converge,  it can potentially speed up the calculations. It also converges rapidly for
the fainter parts of the source, hence it is very useful for sources with high emissivity contrast. One disadvantage is that due to computing resources limitations
it limits the calculations
to only a small number of pre-chosen directions of observations. In addition, for complicated geometries such as those produced in simulations
one must verify that running more photons is  more expensive than calculating $\tau_{esc}$ used in this method.
We find that indeed this is the case for the ART environments where the RT code is applied in this study.

\subsubsection{Parallelization}
To reach high performance we implement the parallel execution of the code. Our Monte Carlo scheme is particularly easy to parallelize, since
each ray is independent of others. The parallelization is done using the 
Message Passing Interface (MPI) library of routines. As every photon ray is independent, 
communication requirements among  the different processes are minimal, and in essence MPI distributes copies of the code which are
run autonomously in the different nodes used. However, each processor is assigned and runs photons from different emission regions.
To get an idea about the performance of the code (using the above acceleration schemes), $10^{7}$ photons 
\footnote{This  number of photons is well above the minimum necessary for the
results to converge as will be discussed in a later section}  transfer to 10 physical kpc from the center of the
ART \lya emitter we  apply the code to in about 
4 hours on 8 Intel Xeon 3.2 GHz processors on the Tungsten NCSA cluster .   

\subsection{Final  images and spectra of simulated \lya emitters}
\label{images_spectra}
The detailed \lya RT is carried out up to a certain distance from the center of the source and then the \lya GP absorption is added.
This distance where the detailed RT stops is determined through a convergence test. 
The existence of such a scale is guaranteed given that the further away a photon moves from the center of the object, the most improbable it becomes for
it to scatter back in the direction of observation.
Furthermore, the size of this convergence radius can also be motivated observationally, from the extent of \lya halos that have been observed.

The surface brightness of each pixel of the constructed image is
\begin{equation}
SB_{p}=\frac{\Sigma_{i,j} F_{i,j} e^{-\tau_{esc,i,j}} P(\phi,\mu)}{\Omega_{pix}} \times e^{-\tau_{GP}} \, ,
\label{sb}
\end{equation}
where the sum is over  the fluxes of all photons ($i$), and all their scatterings ($j$) with scattering positions that project onto the
pixel; $\Omega_{pix}$ is the angle subtended by the pixel to the observer, and the factor $e^{-\tau_{GP}}$ accounts for the
diminishing of the brightness due to the hydrogen intervening between the radius where the detailed RT stops and the observer.  To find the flux $F_{i,j}$ carried by each photon at each interaction, we first calculate the total luminosity, $L_{tot}$, of the
emitter through the sum of the luminosities of the individual source cells.
For $N$ photons (or more accurately wavepackets) used in the Monte Carlo, then each photon carries a luminosity $F_{i,j}$ (independent of photon and scattering numbers $i$ and $j$, respectively,
in our case) equal to
\begin{equation}
F_{i,j}=\frac{L_{tot}}{N} \frac{1}{d_{L}^{2}}
\end{equation}
where $d_{L}$ is the luminosity distance calculated for the adopted cosmology. Note that there is no $1/ 4 \pi$ factor. This factor comes from $P(\phi, \mu)$ -- in equation (\ref{sb}) -- which is
normalized to unity.
 
The GP absorption  optical depth is calculated as described in \citet{hui_etal97}. It is calculated for each pixel separately, and the number of different lines
of sight that have to be used per pixel is determined by checking convergence of the final result. For high enough image spatial resolution (similar to the one used
in this study) one line of sight per pixel is enough, since the simulations themselves have finite spatial resolution.  The 
characteristics of the line emerging after the detailed RT (i.e., its width) and before
adding the GP absorption  determine how far away in distance one must go when calculating $\tau_{GP}$, since one needs to go up to the point where the shortest 
line wavelength
is redshifted at least to the \lya resonance because of Hubble expansion. Often, this physical distance is larger than 
the physical size of the cosmological simulation box. In this case,
we take advantage of the periodic boundary conditions and use replicas of the same box making sure we do not 
go through the same structures. This turns out to be easily done as long as one does not have to use the box
too many times (more than $\sim$ 5). 
Furthermore, we consider two distinct  scenarios, one where the  effect of the red damping wing is taken into account and one where the 
red damping wing is suppressed as would be the case if for example the \lya emitter was in the vicinity of a bright quasar.      

Lastly, spectra are  obtained by collapsing the 3-D image array (2 spatial dimensions+wavelength)  along the spatial dimensions.
 
\section{Application to Cosmological Simulations}
\label{app_sims}
\subsection{The simulations}
\label{sec:sims}
Here we  present some basic information regarding the cosmological simulations we use in what follows in order 
to apply the \lya RT code in a cosmological setting.

The RT is carried out using outputs of  the ART  
code for the concordance flat
{\LCDM} model: $\Omega_0=1-\Omega_{\Lambda}=0.3$, $h=0.7$, where
$\Omega_0$ and $\Omega_{\Lambda}$ are the present-day matter and
vacuum densities, and $h$ is the dimensionless Hubble constant defined
as $H_0\equiv 100h{\ }{\rm km\ s^{-1}\,Mpc^{-1}}$. For the power spectrum normalization
the value $\sigma_{8}=0.9$ is used.
This model is
consistent with recent observational constraints
\citep[e.g.,][]{spergel_etal03}.
The initial conditions of these simulations  
are the same as those in \citet{kravtsov03} and \citet{kravtsov_gnedin05}, leading to the formation of a Milky Way sized galaxy at $z=0$.
However, these simulations are different in that, in addition to 
dark matter, gas dynamics, star formation and feedback, cooling, etc.,  they also include  non-equilibrium ionization and 
thermal balance of H, He, H$_{2}$ and
primordial chemistry,  full RT of ionizing radiation and optically thin line RT of
Lyman-Werner radiation. The continuum RT is 
modeled according to the Optically Thin Variable Eddington Tensor approximation described in \cite{gnedin_abel01},
whereas cooling uses the abundances of species from the reaction network, as well as corrections for cooling enhancement due to metals. 

The code reaches
high force resolution by refining all high-density regions with an
automated refinement algorithm. The criterion for refinement is the
mass of dark matter particles and gas per cell. 
Overall there are 9 refinement levels. The  physical size of a cell of refinement level
$l$ is $26.161 \times 2^{9-l}$ pcs at $z \simeq 8$ (the redshift  we  focus on  in this study). 
The dark matter particle mass at the highest resolution region is $9.18 \times 10^{5} h^{-1} \rm{M}_{\odot}$, and the box size for 
which results are presented in this paper is $6 h^{-1} \rm{Mpc}$.

For each simulation cell we have available information 
such as the 
temperature, the peculiar velocity, the neutral hydrogen density, the ionized hydrogen density, the metallicity, etc.
With this information and using the mesh of the ART code itself we  follow how \lya photons are being initially emitted and subsequently getting
scattered. As an example of an application of the \lya RT code developed for the ART code we  focus on the most massive emitter
at $z \simeq 8$. This emitter is found within a highly ionized, butterfly--shaped bubble. Outside this bubble the Universe is highly neutral, whereas some
dense neutral cores associated with the forming galaxy exist within the bubble. Results for more emitters, different redshifts, multiple directions of observation, larger simulation boxes, etc.,  
will be presented in future papers.
  
\subsection{Intrinsic \lya emission}
\label{intrinsic_emission}
There are a number of different mechanisms that can produce \lya emission from high-redshift objects.
Here we classify them into recombination and collisional emission mechanisms.
By recombination emission mechanisms we refer to \lya photons that are the final result of the cascading of 
recombination photons produced in {\it ionized} gas.
The gas may be ionized by the UV radiation  of hot, young, massive stars, from an AGN hosted by the galaxy, or 
by the  intergalactic UV background. By collisional emission mechanisms we refer to photons
that are produced by the radiative decay of excited bound ({\it neutral}) hydrogen states, with 
collisions being the mechanism by which these excited states are being populated. This mechanism takes place  when
gas within a dark matter halo  is cooling and collapsing to form a galaxy and radiates some of the gravitational collapse 
energy by collisionally excited \lya emission, when  gas is shock heated by galactic winds or by jets in radio galaxies, and 
in supernova remnant cooling shells. We underscore the
fact that the states are bound states, because in principle collisions can also cause ionization in which case we would have production
of \lya photons under a recombination mechanism, according to our definition conventions. 
With the exception of AGN and jets, which are not included in ART  simulations, as 
well as the fluorescence emission due to the intergalactic UV background which would be relevant at lower redshifts than
we focus on in this study, we will try to briefly assess the importance of these separate \lya emission sources. This is  interesting in particular 
because, in addition to the different dependence on the physical parameters (i.e., different temperature dependence and dependence on
ionized versus neutral hydrogen), these mechanisms may also have a different spatial distribution. For example, 
shock heated gas from gravitational collapse may be a spatially more extended \lya source than the gas photoionized 
by UV  radiation of young stars at the relatively
compact star forming regions. The dominant source of \lya emission may be what distinguishes most \lya emitters from the more extended sources 
referred to in literature as \lya blobs \citep{steidel_etal00,haiman_etal00,fardal_etal01,bower_etal04}. 

Before discussing the different  \lya emission  mechanisms,  we should first mention that,
due to practical limitations (i.e., we can only use a relatively limited  number of photons), 
we use as source cells only the cells that contribute significantly to the total luminosity of the object.
Hence, we set a threshold on the cell luminosity and use as source cells only the cells whose luminosity exceeds this
threshold. Then by performing a convergence test, namely by doing runs assuming different luminosity thresholds up to the point where including
lower luminosity source cells does not change the results (within some pre-specified tolerance), we determine the minimum luminosity a simulation
cell must emit to be one of the cells where photons will originate from.
It is meaningful to consider a similar convergence check with respect to the \lya RT results, and this 
will be discussed in a later section.
The convergence test reveals that the luminosity of the object is dominated by a few very luminous cells.
To get an idea,  the luminosities of cells within the virial extent roughly range from  $10^{41}$ to several  times 
$10^{54}$ photons/s.
The total luminosity of the object is the sum of the luminosities of the cells considered. Even though most of the volume, say, within the virial radius
is in low to moderate luminosity cells, the sum of the luminosities of these cells is not significant enough compared to the less numerous
high luminosity  cells. For the object at hand the convergence test suggests that one can use as source cells only cells with 
luminosities above $\simeq 5 \times 10^{50}$ photons s$^{-1}$. This value  determines the relative importance of the different \lya emission mechanisms discussed
in what follows. With the aforementioned luminosity threshold, the total  luminosity of the emitter at hand  is roughly equal to
$4.8 \times 10^{43}$ ergs/s. 
We sample the emission region (i.e., the cells with luminosity above the luminosity threshold discussed) 
by emitting equal weight wave packets, but in numbers 
that reflect the relative luminosities of the  cells.  

Note that this discussion on the various mechanisms, emission rates, etc., should somehow be affected by the limited simulation resolution, a factor
that will be studied in detail in the future.
Furthermore, the approach adopted in this section is  an 'order-of-magnitude' one. We defer a more thorough 
and statistical analysis of the \lya emission sources in high
redshift galaxies to a future study,  where all factors will be taken into account. For example, the discussion about the importance of the various
emission mechanisms  must be extended to the after RT results and after including dust. This 
is because   it could, for example, be the case that recombination \lya photons, despite being more numerous as discussed
below,
may be more likely to be absorbed than collisional \lya photons, if one assumes that there is more dust in star forming  regions -- where recombination
photons are generated -- than in regions where collisional \lya photons originate from. 

\subsubsection{\lya photons from recombinations} 
The recombination rate of a cell is
\begin{equation}
r=n_{e}n_{p} \alpha_{B}V
\label{rec_rate}
\end{equation}
with $n_{e}, n_{p}$ the number density of electrons and protons, respectively, and 
$V$ the volume.
In principle species other than hydrogen may  contribute to $n_{e}$. Thus,  $n_{e}$ in general is not equal to $n_{p}$.
In what follows, we take into account electrons contributed by the ionization of He.
Other BBN predicted species such as Li, Be and B  
(with, anyway, tiny abundances), and  elements produced through stellar processing such as C, N and O  
are not taken into account.

Recombination photons are 
converted with certain efficiency into \lya photons. 
In particular,  for a broad range of temperatures centered on $T=10^{4}$ K, roughly
$38 \%$ of recombinations go directly to the ground state. 
A fraction $\sim 1/3$ ($32 \%$) of the recombinations that do not
go to the ground state go to $2S$ rather than $2P$ and then go to the ground state via two continuum photon decay \cite[cf. Table 9.1 of][]{spitzer78}.
Hence, only a fraction
$\sim 40 \%$ of the recombinations yield a \lya photon. The temperatures of simulation cells within the virial extent
of the emitter are in the $10^{2.4}-10^{6.3}$ K range, with most cells in the $10^{4}-10^{6}$ K range. Due to the weak temperature dependence of the 
various recombination coefficients  the above
conversion efficiencies are roughly applicable throughout this temperature range. Furthermore, if the gas is optically thick, then
photons that originate from recombinations  to the ground state will be immediately absorbed by another
neutral hydrogen atom and eventually they, as well, will produce \lya photons. 
Assuming for now that this is the case (as will be discussed  later in this section), as well as that  the medium is thick in Lyman-series photons, so that 
all higher Lyman-series photons are
re-captured and eventually yield \lya photons, we  adopt   case B recombination. For 
the recombination coefficient we use the fit obtained by  \citet{hui_gnedin97}, accurate to $0.7\%$ for temperatures from 1 to
$10^{9}$K
\begin{equation}
\alpha_{B}=2.753 \times 10^{-14} \rm{cm}^{3} \rm{s}^{-1} \frac{\lambda^{1.5}}{\left[1+\left(\frac{\lambda}{2.74}\right)^{0.407}
\right]^{2.242}}
\end{equation}
with $\lambda=2 T_{i}/T$, and $T_{i}=157807$ K the hydrogen ionization threshold temperature.
In agreement with the above argument, the effective recombination coefficient at level 2P is approximately 
$2/3$ of the case B recombination
coefficient and that is what we use to convert recombination rates into \lya photon emission rates. 
Thus we assume that the conversion  efficiency from recombination to \lya photons is exactly the same for all simulation cells.
This is a good assumption since the conversion efficiency has a very weak temperature dependence.

The exact conversion efficiency for each source cell  also depends  on the rate at which collisions redistribute atoms between
the 2S and 2P state. 
Collisions with both electrons and protons are relevant. 
To get an idea for the cross sections involved, for
a temperature of $10^{4}$K and thermal protons $\sigma_{2S\rightarrow 2P} \simeq 3 \times 10^{-10} \rm{cm}^{2}$ \citep{osterbrock89}.
For thermal protons and electrons the thermally averaged  collisional cross sections for
the processes
\begin{equation}
H(2P)+p\rightarrow H(2S)+p
\end{equation}
and
\begin{equation}
H(2P)+e\rightarrow H(2S)+e
\end{equation}
are  $q_{p}=4.74\times 10^{-4} \rm{cm}^{3}/\rm{s}$ and $q_{e}=5.70\times 10^{-5} \rm{cm}^{3}/\rm{s}$, respectively,
for a temperature of $10^{4}$K \cite[cf. table 4.10 of][]{osterbrock89}.
The $2P$ to $2S$ transition is relatively important when the proton number densities are  small ($<10^{4} \rm{cm}^{-3}$), and
in this case there is some probability that the \lya photon  gets destroyed through a two quantum decay. For higher densities the opposite conversion
($2S$ to $2P$)  becomes important, canceling out the destruction effect \citep{osterbrock89}.
At the lower density regime, which is applicable in the simulations since there
$n_{p}<10^{4}$ cm$^{-3}$ everywhere (within the virial extent the proton number density range is $10^{-4}-10^{2.5}$ cm$^{-3}$, 
with most cells in the range $10^{-3}-1$ cm$^{-3}$), we can check how important this process really is by
comparing the radiative decay time and the typical time between collisions,
\begin{eqnarray}
\nonumber
p=\frac{q_{p}(T)n_{p}+q_{e}(T)n_{e}}{A_{21}} \\ 
\simeq 8.5 \times 10^{-13} n_{p} T_{4}^{-0.17}
\label{collisions_prob}
\end{eqnarray} 
where the number densities of protons and electrons were assumed to be roughly the same and in $\rm{cm}^{-3}$, $A_{21}=6.25\times 10^{8} \rm{s}^{-1}$ 
is the spontaneous radiative decay for the \lya transition, and  temperature is measured in $10^{4}$K units. The temperature 
dependence of the collision rates is taken from
\cite{neufeld90}.
For the temperature and proton/electron density ranges relevant to the source cell conditions, the probability for a collisional 
$2P$ to $2S$ transition is negligible, at least for the initial emissivity.
We  discuss 
 their effect during scattering of the photons in \S \ref{sec:collisions}.
\begin{figure*}[thb]
\centerline{{\epsfxsize=3.5truein\epsffile{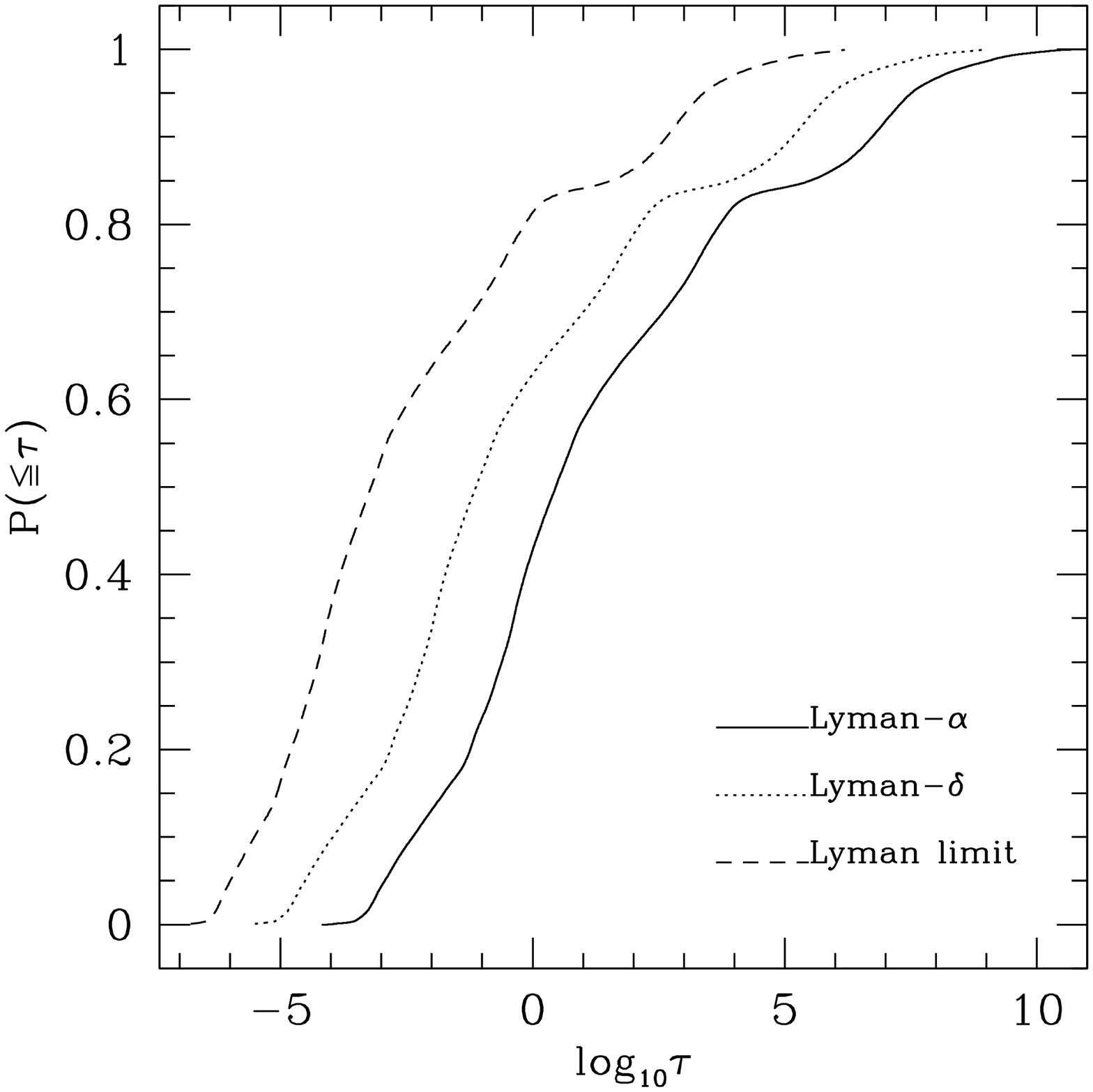}}\hspace{0.5cm}{\epsfxsize=3.5truein\epsffile{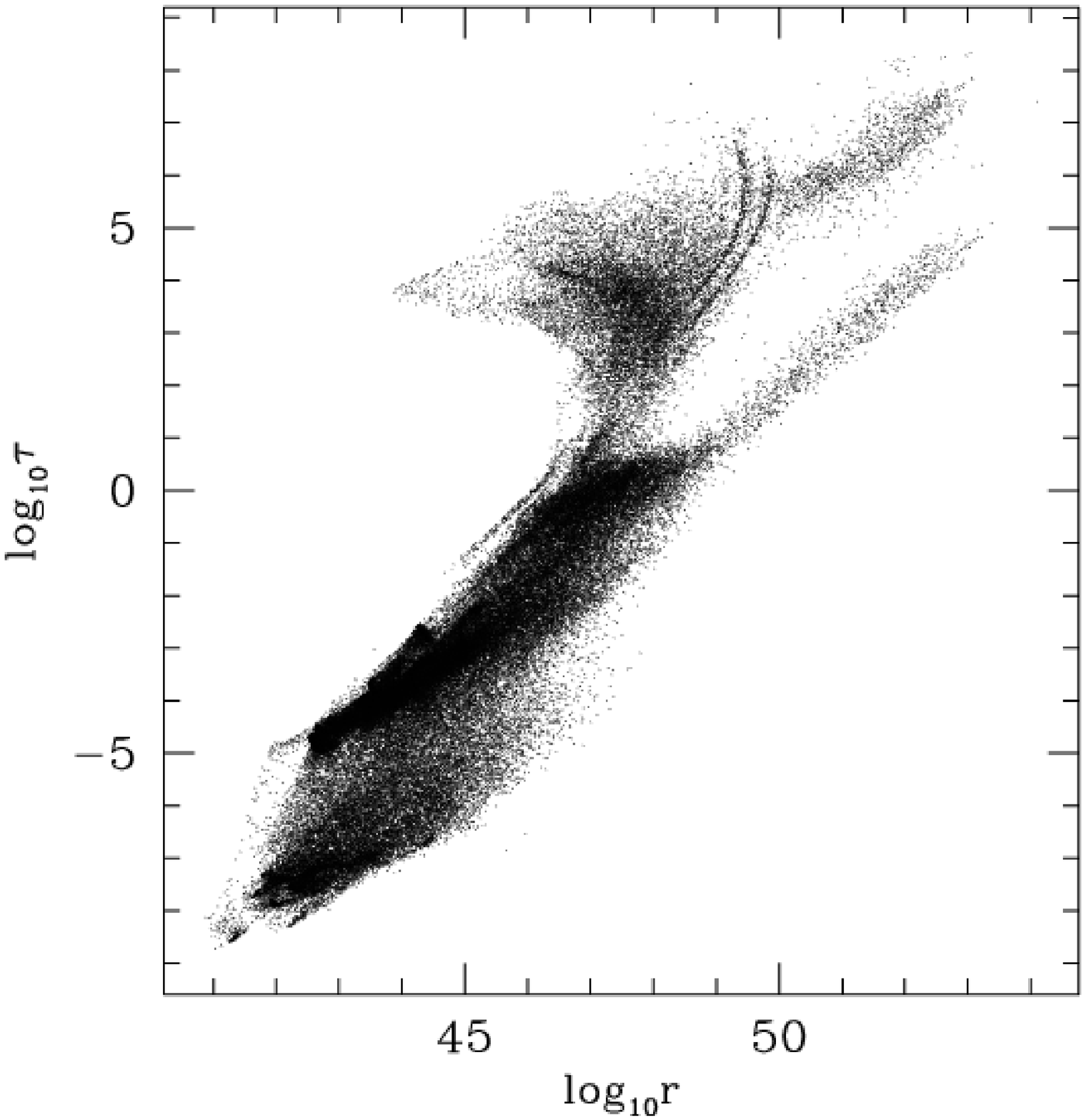}}}
\caption{{\it Left panel:} Cumulative probability distribution of center-of-line optical depths of ART 
simulation cells within the virial extent.  The three different lines correspond
to the cell optical depth distribution in \lya ({\it solid}), Ly-$\delta$ ({\it dotted}) and Ly-limit ({\it dashed}) photons.  
{\it Right panel:} \lya center-of-line optical depth of  the simulation cells within the virial extent of the emitter plotted against the cell 
recombination rate. Since only cells with the highest recombination rates ($\ge 10^{51}$ s$^{-1}$ or, equivalently, luminosities roughly
$\ge 5 \times 10^{50}$ photons s$^{-1}$) need to be used as source cells,  and almost all of 
these cells
have $\tau \ge 10^{3}$, roughly speaking our 'on-the-spot' approximation is satisfactory (see text for details).
\label{lyman_series}}
\end{figure*}

One assumption that we make is that the  cascading of the Lyman series  photons, as well as the re-emission and re-absorption of photons
from recombination to the ground state,  is done  'on-the-spot', namely, locally. In our case ''locally''  means within the same simulation cell.
This assumption  is essential if one wants the \lya emissivity of a cell to depend on its own recombination rate only. If not, one 
faces the complicated situation
where the \lya emissivity of one cell depends on the recombination rates and photon cascade processes that are happening in other cells as well.
The validity of our assumption depends on the optical depth of  Lyman series and ionizing photons  when traversing 
a typical cell in the simulation (and 
should also be affected somewhat by resolution). 
In the left panel of Figure \ref{lyman_series} we show the optical depth  probability distribution function for Ly-$\alpha$, Ly-$\delta$ and Ly-limit radiation. 
The distribution function has as independent variable the optical depth of simulation cells  within  10 physical kpc ($\simeq$ virial extent) 
from the
center of the emitter.
These distributions are very similar, differing only by the values of $\tau$ because of different oscillator strengths and 
characteristic frequencies.
Clearly, in all cases more than half potential source cells are not optically thick, and this is expected to get worse for ionizing radiation beyond the
Lyman limit. However, as shown in the right panel of Figure \ref{lyman_series} the optical depth of a cell correlates 
with its recombination rate. In this figure the optical depth plotted
is that for \lya photons, but it is easy to see how this scales approximately with optical thickness for other Lyman-series photons. Since 
only cells with recombination
rates higher than $10^{51}$ s$^{-1}$ (or equivalently with luminosities higher than roughly $5 \times 10^{50}$ photons s$^{-1}$)  
are used as source cells, our 'on-the-spot' 
assumption seems pretty satisfactory, if not always
accurate. It becomes less and less accurate the higher we go in the Lyman series, and of course beyond the Lyman limit but for the time being
we content ourselves with this approximation, given the complexities introduced when this assumption is not adopted. We will investigate this point further
in the future. 

Lastly, to get an idea about the physical conditions of the highest recombination rate (luminosity)  source cells, they 
 consist of two classes with respect to temperature and
neutral hydrogen fraction: one class contains cold gas elements ($T \sim 10^{3}$K), 
with a neutral hydrogen fraction $>0.9$ (and high gas number density). The second class of very luminous cells consist of warmer
gas elements ($T\sim 10^{4}$ K and a bit higher). 
In the context of \lya cooling radiation, discussed in the next section, 
the first class of cells are unable to cool via atomic hydrogen cooling  since they are cold, whereas the
second class of  most luminous cells could  cool via atomic hydrogen cooling temperature-wise, but that is not happening
because these cells are highly ionized. 
\begin{figure}[hbt]
\centerline{{\epsfxsize=3.5truein\epsffile{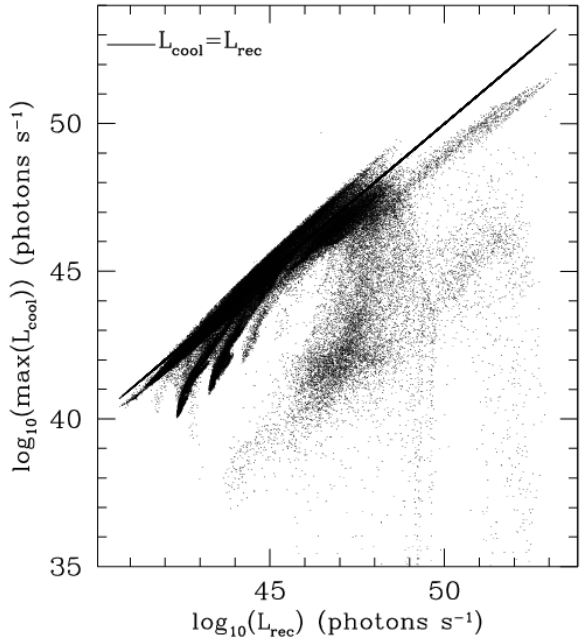}}}
\caption{Maximum cooling \lya luminosity, $L_{cool}$,  plotted against recombination \lya luminosity, $L_{rec}$, for 
all ART simulation cells within the virial extent of a \lya emitter at
$z\simeq 8$. The cooling luminosity is the maximum possible \lya luminosity from cooling because it is derived assuming that all cooling radiation is
emitted in \lya photons. The solid line shows the case where the two luminosities are equal. Since, as discussed in the text, only cells with luminosities 
roughly above $5 \times 10^{50}$ photons s$^{-1}$ contribute significantly to the
luminosity of the emitter,  this figure shows  that recombination is dominant
over cooling \lya radiation. 
\label{cooling_vs_recombination}}
\end{figure}

\subsubsection{\lya photons from collisional excitations}
A collisional emission mechanism whose importance for the simulated objects 
can be assessed relatively easily is that of atomic hydrogen cooling. Using the expression by \citet{hui_gnedin97} 
for the hydrogen cooling rate (used in the ART 
simulations analyzed here), 
and assuming for the moment that this energy is all emitted in the form of \lya photons, we obtain for the luminosity (number of \lya photons/s) emitted by a cell
\begin{equation}
L_{cool}=4.6 \times 10^{-8} \frac{e^{-1.18355/T_{5}}}{1+T_{5}^{0.5}} n_{e} n_{HI} V \, 
\end{equation}
with $T_{5}$ the temperature in units of $10^{5}$ K. 
This is compared with the recombination luminosity $L_{rec}$ ($\simeq 0.68 r$) in Figure \ref{cooling_vs_recombination}. 
Taking into account the results of the convergence test performed to specify what is the minimum cell luminosity that needs to be taken into account
($\sim 5 \times 10^{50}$ s$^{-1}$), we see that 
the cells which are relevant are 
cells  where recombination processes dominate, as can be seen in Figure \ref{cooling_vs_recombination}. Namely, 
similar to previous studies \citep[e.g.,][]{fardal_etal01} we find that 
the cooling radiation \lya contribution is subdominant compared to the recombination contribution, hence in what follows we focus only on the latter.   

\subsubsection{Supernovae Remnants (SNR)}
A \lya source that  yields \lya photons both from recombinations and collisional excitations is supernova remnants (SNR).
\citet{shull_silk79} have computed the time-averaged \lya luminosity of a population of Type II SNR using a radiative-shock code.
They find that the \lya luminosity of a galaxy due to  SNR
is $L_{SNR}=3\times10^{43} n_{H}^{-0.5} E_{0}^{0.75} \dot{N}_{SN}$ ergs/s,
with $n_{H}$  the ambient density in cm$^{-3}$, $E_{0}$ the typical supernova energy in units of $10^{51}$, and $\dot{N}_{SN}$ the number of
supernova per year. Strictly speaking, this quantity also depends on the assumptions on the IMF, and the lower and  upper stellar masses of the 
mass range over which the IMF is to be integrated. This expression includes both contributions, from  recombination and collisional emission
mechanisms: from UV and X-ray ionization (coming from the hot SNR interior) of the surrounding medium  and from
cooling shells, respectively. A thorough investigation of the relative importance of SNR \lya emission with respect to
that from young stars photoionization has been carried out by \citet{charlot_fall91,charlot_fall93}. The general conclusion reached is that for
a broad range of physical conditions and assumptions, the SNR contribution is at best a factor of 2.5 less than that from stellar ionizing
radiation.  
These results make the effort to include the (anyway not resolved in ART simulations) SNR contributions  superfluous.

\subsection{The \lya emitter before  RT}
\label{beforert}
\begin{figure*}[bht]
\centerline{{\epsfxsize3.5truein\epsffile{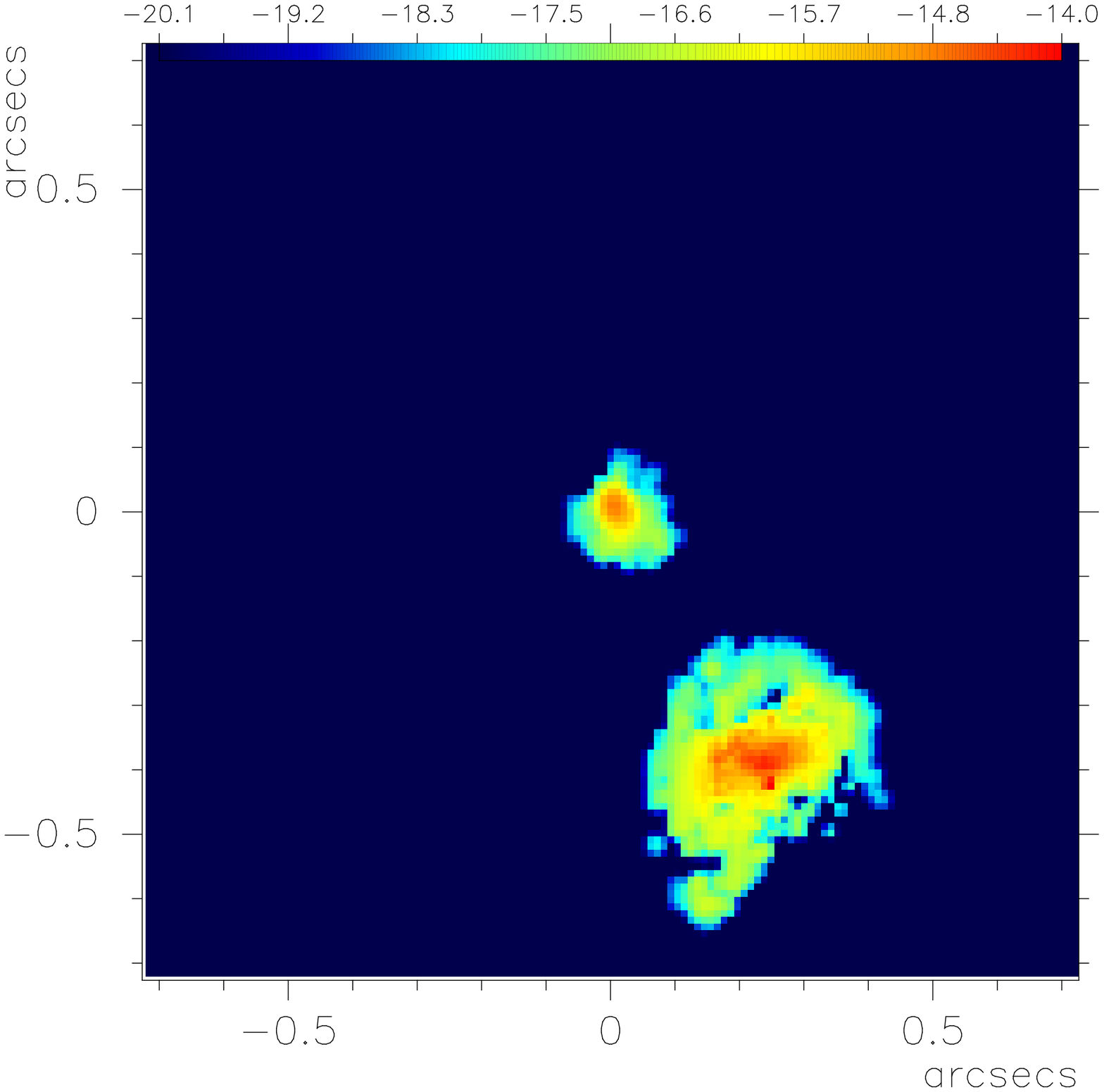}}\hspace{0.5cm}{\epsfxsize3.5truein\epsffile{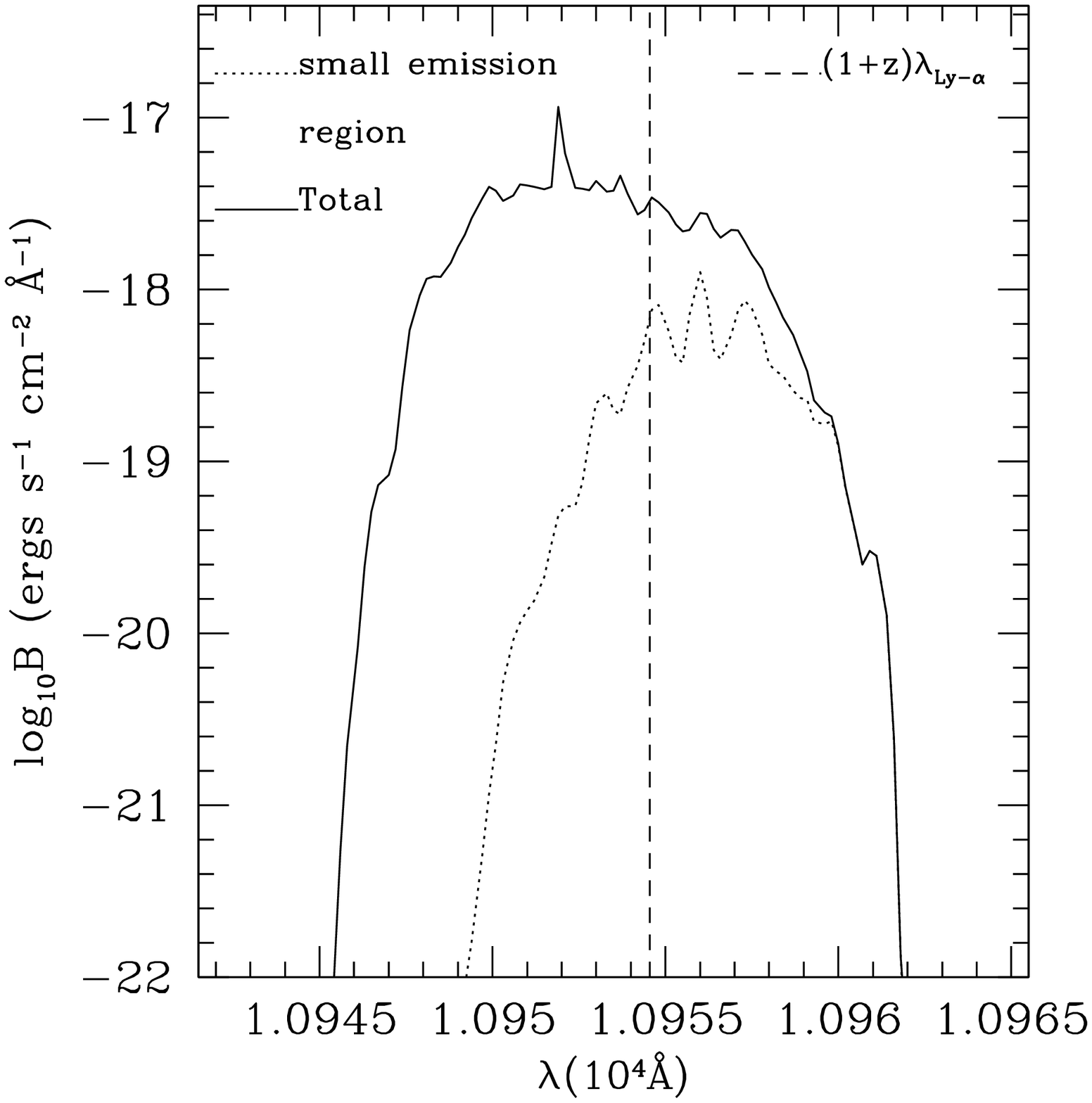}}}
\caption{{\it Left panel:} Image  of \lya direct emission (i.e., assuming the \lya photons escape directly to the observer after they are produced). The
approximately $1.4 \times 1.4$ arcsecs$^{2}$ ($\simeq 6.5 \times 6.5$ physical kpc) field corresponds to roughly one third of the virial extent of the dark matter halo where the emitter lives. 
The surface brightness (SB) is bolometric and in units
of $\rm{ergs} \ \rm{s}^{-1} \rm{cm}^{-2} \rm{arcsecs}^{-2}$. The SB color scale is logarithmic. The object had undergone a recent merger, that is why there 
are two distinct luminous blobs that dominate the emission. {\it Right panel:} Initial \lya injection spectrum. Shown are the total spectrum 
({\it solid line}), namely the spectrum for the image shown in the left panel, and the spectrum of the smallest of the two blobs in the image ({\it dotted line}).
Note that the wavelength is in $10^{4} \AA$ (i.e., $\mu m$). The dashed line shows the \lya resonance for $z \simeq 8$. See text for discussion of the structure of the line.  
\label{fig:iniemission}}
\end{figure*}

To get an idea of the size of the emitting region, the prevailing physical conditions, and for comparison with results obtained later after including 
RT, in this subsection we briefly present the emission spectrum and image of the emitter as they would appear to an observer at $z=0$ if the \lya photons
escaped  without any scattering.
An image and a spectrum 
of the emitter along a certain direction of observation is shown in the left and right panel, respectively,  
of Figure \ref{fig:iniemission}.

The image is a surface brightness map (in units of ergs s$^{-1}$ cm$^{-2}$ arcsec$^{-2}$) of a roughly 
$1.4 \times 1.4$ arcsecs$^{2}$ field which corresponds to approximately one third of the virial extent of the dark matter halo the emitter lives in
(with the virial extent $\simeq20$ physical kpc in diameter). There are two distinct emission regions, each one corresponding to the two progenitors that merged and formed this object. 
The color scale for the surface brightness is logarithmic. Clearly, the emission region is very small (the largest of the two structures is 
at most $\simeq 2-2.5$ physical kpc in diameter, if one includes  the faintest pixels), compared for example to the virial extent of the
dark halo. The resolution of this image is $0.01$ arcsecs ($\simeq$ 0.05 physical kpc), at least 10 times higher than the best resolution currently available. 
As discussed before,  for these results only cells with recombination rates higher or equal to $10^{51}$ s$^{-1}$ are used. 
The initial frequency is chosen according to  a Voigt profile that is sampled for each cell out to $10$ Doppler widths and shifted around the bulk
(peculiar + Hubble) velocity component along the direction of observation.  
 The number of photons used ($3 \times 10^{5}$) has been determined after a convergence study. 
Note that when we study the convergence with respect to the number of photons we take into account that this must be done
in parallel with how far away in the wings
we go when sampling the emission Voigt profile of each cell, since the higher the number of
photons used the better one can sample frequencies further away from resonance. The convergence procedure gave the aforementioned number of photons and
initial emission frequency range (i.e., 10 thermal Doppler widths). 
 
In the right panel of Figure \ref{fig:iniemission}, 
 the frequency resolution  is $\lambda/\Delta \lambda \sim 50000$. 
The 
 line shape has converged, namely the peaks shown correspond to real velocity
 substructure. For example, the most pronounced peak at $\lambda = 10952 \AA$ corresponds to the component of the peculiar velocity
 along the direction of observation of the most luminous pixel of the image shown at the left panel (with coordinates on the image
 (0.24,-0.42) arcsecs, roughly). The 
 dominant contribution to this 
 pixel comes from the highest recombination cell of the emitter with a recombination rate equal to $\simeq 1.3 \times 10^{55}$ s$^{-1}$ and 
 a peculiar velocity component along the direction of observation equal to $0.27 \times 10^{-3}$ the speed of light.  
 As mentioned,  for each emission cell the Voigt profile was used and sampled up to 10 thermal Doppler widths. 
 The total width of the line however is dominated by the bulk velocity structure of the emitter. The full width of the line
 at the minimum flux level shown in the figure ($10^{-22}$ ergs $s^{-1}$ cm$^{-2}$ $\AA^{-1}$) is  roughly $15 \AA$ (with the width
 if bulk velocities are set to zero being less than half this). This width corresponds to 
 projected velocities along the direction of observation roughly in the $[-200,200]$ km/s range (this is just approximate, 
  note however that the line is not symmetric around the
 rest frame resonance). This velocity range is what is expected given the peculiar velocities of the emitting cells. 
Also shown is the  spectrum of the smallest of the two substructures ({\it dotted line}) of the  image shown in the left panel. One can 
 easily infer what the spectrum of the large \lya substructure looks like.

The results discussed in this section may be  specific to the emitter at hand, but the considerations themselves are pretty general. 
The same kind of procedure must be repeated for each individual emitter identified in the simulations.

\subsection{The \lya emitter after RT}
\label{afterrt}
\begin{figure*}[th]
\centerline{{\epsfxsize3.5truein\epsffile{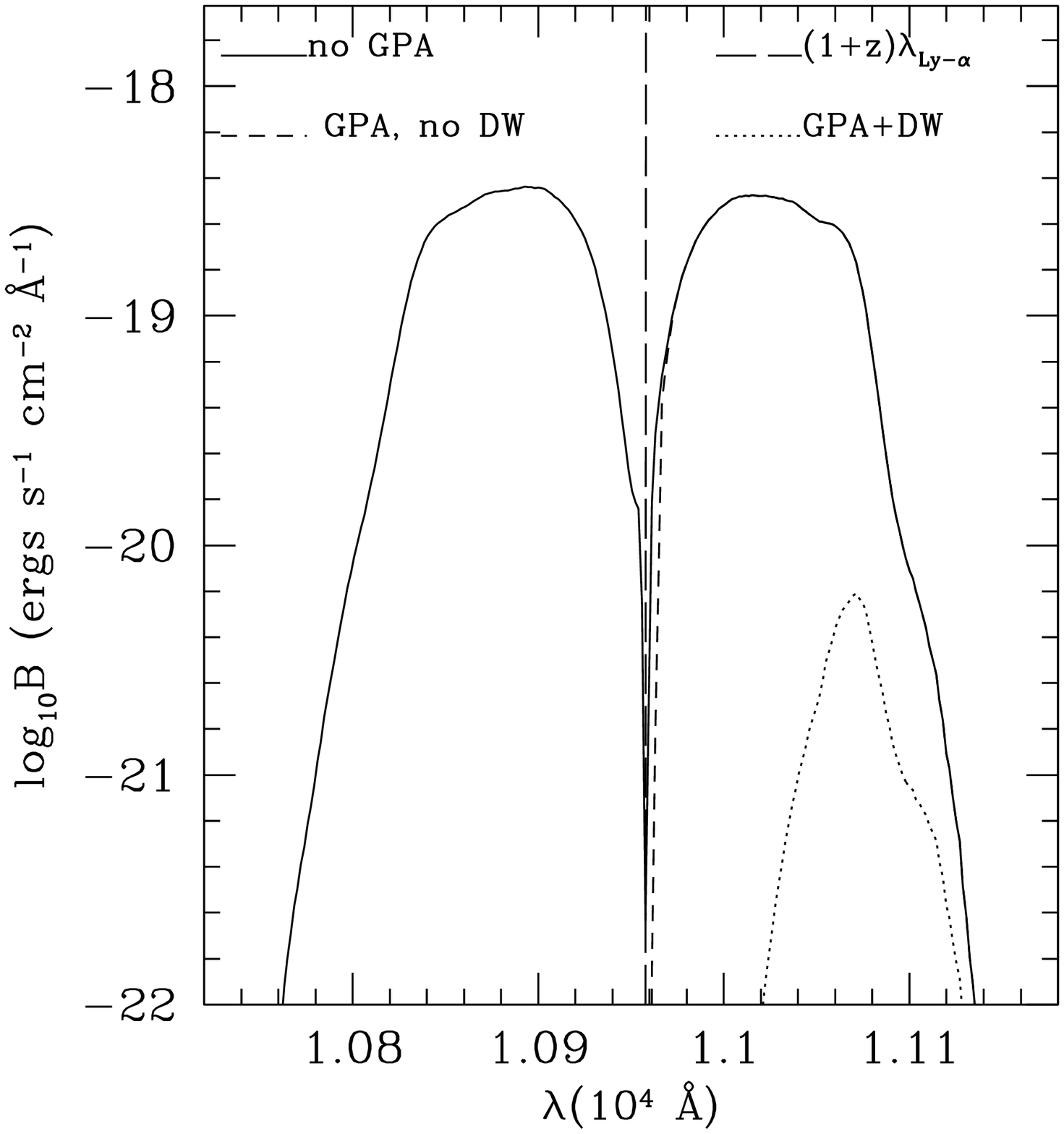}}\hspace{0.5cm}{\epsfxsize3.5truein\epsffile{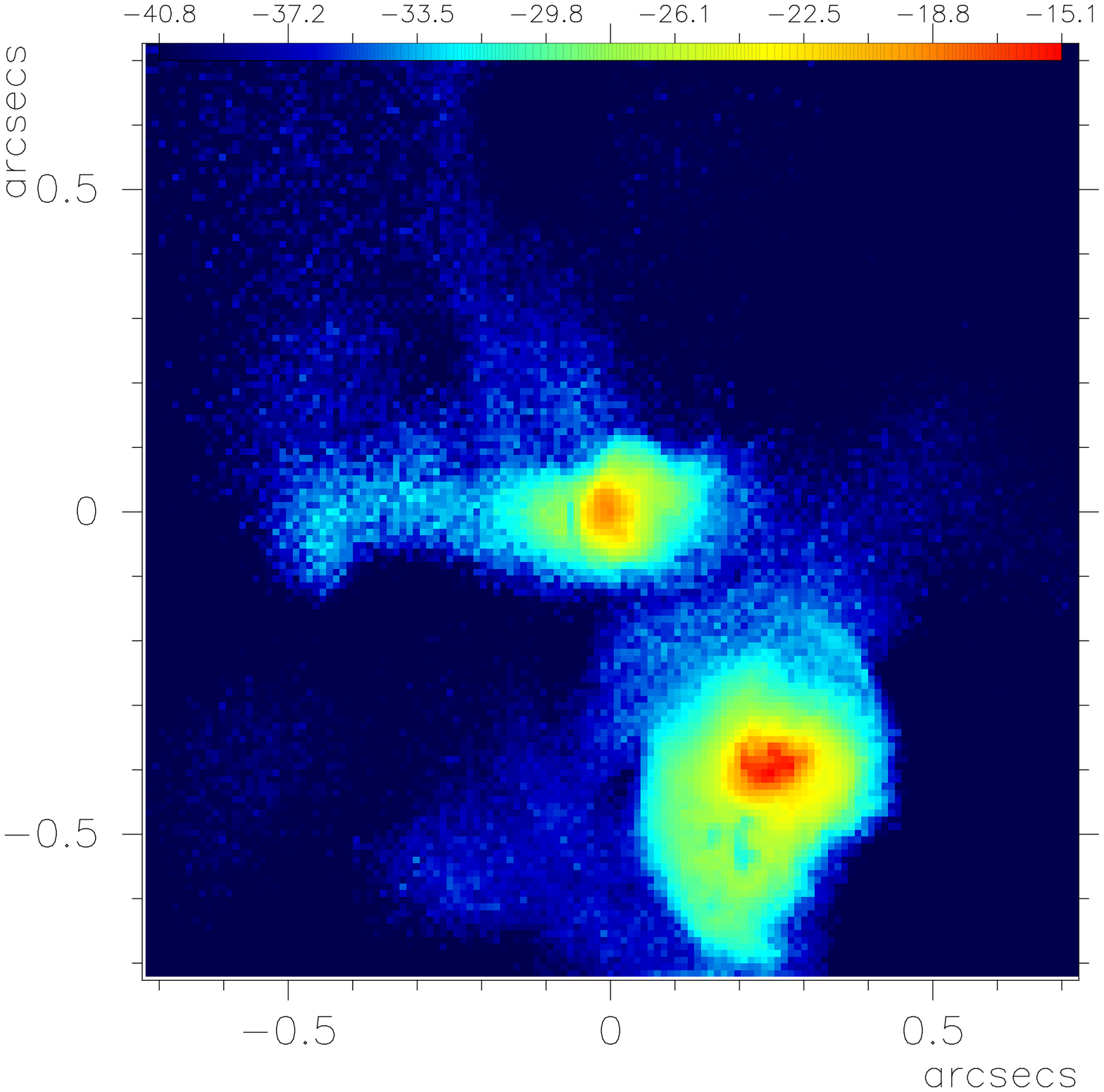}}}
\caption{{\it Left panel:} Emerging \lya emission spectrum before adding the \lya Gunn-Peterson absorption (GPA) 
({\it no GPA, solid line}), with the GPA  but without the red damping wing ({\it GP, no DW, short-dashed line}), 
and with GPA and the damping wing ({\it GPA+DW, dotted line}). 
Note that the wavelength is in $10^{4} \AA$ (i.e., $\mu m$). The long-dashed line shows the \lya resonance for $z \simeq 8$.
 {\it Right panel:} Image  of the \lya emitter after RT, GPA and DW. The
$1.4 \times 1.4$ arcsecs$^{2}$ ($\simeq 6.5 \times 6.5$ physical kpc) field corresponds to roughly one third of the virial extent of the dark matter halo where the emitter lives. 
The surface brightness (SB) is bolometric and in units
of $\rm{ergs} \ \rm{s}^{-1} \rm{cm}^{-2} \rm{arcsecs}^{-2}$. The SB color scale is logarithmic.    
\label{rt_only}}
\end{figure*}

It is interesting to first treat  the emitter as a finite configuration. In this case,  as soon as the photons exit this configuration (whose size is 
taken to be roughly equal to the virial extent of the object, namely 10 physical kpc)
they travel towards the observer. In other words at first we ignore  the effect of the GP absorption.  This 
context is pretty similar to that of \S \ref{sec:test} and \S \ref{simple_models}.
We focus on the emergent spectrum shown with the solid line in the left panel of Figure \ref{rt_only} . 

The spectrum converges if  $3 \times 10^{5}$ photons are used, namely if the same number of photons are used as the number 
of photons  needed for the initial emission results (discussed in
\S \ref{beforert}) to converge. Of course, the higher 
the number of photons the better one samples low intensity wavelengths.  We find that the number of 
photons used affects wavelength ranges  with  flux less than about $10^{-22}$ 
ergs s$^{-1}$ cm$^{-2}$ $\AA^{-1}$.  The spectrum shown in Figure \ref{rt_only} is produced  using $10^{7}$ photons. 
The spectral resolution used in the figure is  $\lambda/\Delta \lambda \simeq 5000$, whereas  the spectrum is 
identical if ten times
better resolution is used. 
We have performed a large set of convergence tests among which the most interesting are for  different (smaller) cores for the acceleration scheme discussed in
\S \ref{skipping}, and /or  a larger minimum $\tau_{0} \phi(x_{i})$ for which the acceleration scheme discussed in \S \ref{controlled} is used.
Our results are pretty robust, as should following the discussion in \S \ref{controlled} and \ref{skipping}
with respect to the one cell convergence results.

Even though meant for a slab, it is interesting to check if some predictions 
of the Neufeld solution, such as the frequency where the spectrum has a maximum ($\simeq 0.9 (\alpha \tau_{0})^{1/3}$), are roughly
in agreement with the spectrum of the simulated emitter.
Of course, the \lya emitter environment is neither isothermal nor homogeneous, and it is not obvious how to define an 'effective' temperature and
optical depth for these purposes.
Thus, focusing on order of magnitude checks,  setting the expression for  the frequency where the peak  emission occurs in a slab  
equal to the frequency where
the spectrum of the emitter peaks (say in red wavelengths) one finds that the 'effective' optical depth and 'effective' temperature of the equivalent
slab (i.e., the slab that would give a spectrum with peak at the frequencies where the emitter spectrum peaks) roughly satisfy the relation
$\tau_{0} T \simeq 1.4\times 10^{10}$ with $T$  measured in eV. 
The effective optical depth  will be at least 
equal to the most optically thick cell the photon found itself in. Since the emission originates from the most optically thick cells (see 
Figure \ref{lyman_series}), 
$\tau_{0}$ will be at least $10^{3}$. If we assume for example a temperature $T=10^{5}$K, the above relation yields $\tau_{0} \simeq  10^{9}$ which is roughly the optical depth
from the center of the object to its virial radius along the direction of observation.
Thus, the maximum of the spectrum is roughly where it is expected to be if one assumes the scaling from the Neufeld solution ($\simeq 2550$ km/s).      

The emerging spectrum looks pretty similar to the spectrum that would emerge from a static
configuration, namely it has two quite similar peaks, one to the red and one to the blue of the \lya resonance.
Note however that the peaks are not really symmetric, since the flux  decreases more rapidly near the resonance. 
The width of the blue peak at a flux level of $10^{-22}$ ergs s$^{-1}$ cm$^{-2}$ $\AA^{-1}$ is roughly 180 $\AA$ or $\simeq 5000$ km/s. 
We obtain quite a similar spectrum if we set the bulk velocity field to zero in the code, that is kinematics do not seem to play a crucial
role in this case.
In the case of the specific  \lya emitter and for the specific direction of observation, 
analyzing the bulk velocity field (i.e., the peculiar velocity field
since the Hubble expansion is negligible at the distances we are working) we find that there is some net infalling motion, but with
significant transverse velocity components as well. Hence, the obtained static--like spectrum does not come as a surprise.  
Furthermore, the peak asymmetry due to the existence of bulk fields depends on the relative magnitudes of the bulk and thermal velocities (e.g., 
if the bulk velocity is close to the thermal we do not expect a significant asymmetry since one scattering can give, e.g.,  a red photon 
moving in a contracting medium a large enough shift to erase the effect of the contraction) which varies from cell to cell, and it also
depends on the optical thickness. Since thermal velocities are typically small compared to bulk velocities in simulations, 
the optical thickness is a more crucial factor. For such extremely optically thick media where the spectrum is expected to have 
in the context of the Neufeld solution a typical frequency  of $\simeq 2550$ km/s, bulk velocities of at most some hundreds km/s 
will not really favor blue versus red photons (even if the bulk motion was purely inwards) that much, since both red and blue  photons see a very optically thick
medium.

It would be interesting to have a sense of what is the  number of scatterings each photon undergoes before exiting. With 
the acceleration methods that we have to use though it is difficult to
keep track of this quantity. 
A simple way to obtain an order-of-magnitude idea of the number of scatterings in such or, at least,  similar configurations can be obtained  by one of the
examples discussed in \S \ref{simple_models}. For the most optically thick case ($\tau_{0} \simeq 8.3 \times 10^{6}$) and a point source emitting photons that 
propagate in a stationary medium the number of scatterings in one run of $\simeq 2000$ photons varies from $2.5 \times 10^{3}$ up to $4\times 10^{7}$, with an average of
$8.3\times 10^{6}$, and a median of $6.6 \times 10^{6}$. Two thirds of the photons are in the $[4.6 \times 10^{6},2.1 \times 10^{7}]$  scatterings range.
More generally,  we find  that similar to  the Neufeld problem, the average number of scatterings in this spherical configuration 
scales linearly with optical depth at such thick media (see discussion in \S \ref{sec:neufeld}), 
with the proportionality
constant of order unity. From this linear scaling of the average number of scatterings with optical depth,
  one can obtain a rough  idea of the average number of scatterings of photons in the simulation environments (for the cell optical
  depth range in the simulations see, e.g., the left panel of Figure \ref{lyman_series}).
These numbers also make clear  why it is absolutely not feasible to perform \lya RT in the much thicker and more complicated 
simulation environments without some acceleration schemes.

Photons at very optically thick regions have to shift off resonance significantly to escape, and hence
are the ones responsible for the significant line width of the spectrum (along with the $1/x^{2}$ behavior of the wing optical depth, as
discussed previously). It is meaningful to ask whether one should really care about these photons, or instead
ignore them because  may be they are trapped indefinitely (for any practical purpose) in the dense cells and do not participate in the radiation propagation.
To answer this question we estimate the photon diffusion time and compare it to the sound crossing and dynamical time scales (other time scales, such as
the Hubble time scale for example which is $\sim 1$ Gyr at $z=8$  are clearly large enough to be non-relevant). 
Same as with the number of scatterings, to find the exact diffusion times one should follow the detailed RT. Given our acceleration methods this is not
 done. Instead we use some useful scalings. 
Since the average number of scatterings in very optically thick  media is roughly equal to $\tau_{0}$, then the diffusion time
is roughly $t_{d} \simeq N_{sc} l_{mfp}/c$ with $l_{mfp}$ the mean free path between scatterings defined through 
$\langle\tau\rangle=\int_{0}^{\infty} \tau e^{-\tau} d\tau=1$. 
In other words, since $\tau_{0}=n \sigma(x=0) L$, $\tau=n \sigma(\tilde{x}) l$, then the mean free path in units of the total (half) width of the slab is $\sigma(x=0)/\sigma(\tilde{x}) 1/\tau_{0}$,
with $\sigma(x=0)$ the cross section at the line center and $\sigma(\tilde{x})$ the cross section calculated at an effective $\tilde{x}$ so that the above definition for the
mean free path is valid.
Substituting  in the expression for $t_{d}$ we obtain $t_{d} \sim \sigma(x=0)/\sigma(\tilde{x}) L/c$.
For a slab with $\tau_{0}=10^{6}$ we obtain a mean number of scatterings equal to $9.5 \times 10^{5}$ and 
a median equal to $7.2 \times 10^{5}$, whereas $67\%$ of the photons have between $5.1\times 10^{5}$ 
and $2.3\times 10^{6}$ scatterings. For the mean free path we find a mean equal to
$2.4 \times 10^{-5}$, a median equal to $1.9 \times 10^{-6}$ and $67 \%$ of the scatterings correspond to  mean free paths  between $1.4 \times 10^{-6}$ and
$7.7 \times 10^{-6}$, all in units of the (half) width of the slab $L$. 
For the total distance traveled by the photons before escaping, we find an average distance of 40.2, a median of
32.3 -- implying a $\sigma(x=0)/\sigma(\tilde{x})$ ratio of order 10 -- whereas $67\%$ of photons exit after traveling a distance between 16.7 and 96.3, with these numbers as before in units of the width of the
slab.   Based on spatial random walk arguments one would have $N_{sc} \sim \tau_{0}^{2}$, hence the distance before escape would be $\sim \tau_{0}$
or $10^{6}$ for the specific example we use here.  However, as discussed in \S \ref{sec:neufeld} 
$N_{sc}$ scales linearly with $\tau_{0}$ and this makes a big difference. 
We find that the sound crossing time is significantly higher than the dynamical time for most simulation cells, hence the latter is the relevant time against which 
the diffusion time must be compared.     
We find that the dynamical time scale is at least three orders of magnitude or more larger than $L/c$ 
which is within a factor of order $10^{2}$ -- for the various physical conditions in the simulation cells -- representative of $t_{d}$. Note that this comparison also justifies the use
of 'static' simulation outputs where the RT is performed, even though we plan on investigating the possibility of incorporating the
RT scheme into the dynamical evolution in the simulations.
Furthermore, the effect of the simulation resolution on these conclusions will be investigated in a future study. 
 
The scattering process  diffuses the initial number of emitted photons on a larger area and hence lowers the {\it number} surface brightness 
(i.e., number per s cm$^{2}$ arcsec$^{2}$ rather than energy per s cm$^{2}$ arcsec$^{2}$). In general the surface brightness itself can go either up or down, depending
for example on the velocity structure of the medium. To quantify this effect on a photon-by-photon basis 
we choose to calculate the distance on the plane of the image between the initial
emission point and the point (pixel) where the photon makes its maximum contribution to the image (see \S \ref{images_spectra} 
on how spectra and  images are obtained). 
We find that these distances vary from roughly $10^{-3}$ to $10$ physical kpc, with a median of 0.27 kpc and a mean of 0.31 kpc.
Given that the largest of the two  emission regions has a diameter of $\sim 2-2.5$ physical kpc (see, Figure  \ref{fig:iniemission}),
this means that the 'size' of the luminous part of the object increases on average by more that 10\% due to scattering.
If, instead, we focus on the region where a certain fraction of photons originates from we obtain quite similar results. For example, 
ignoring the effects of RT,  $90\%$ of the emitted photons that would reach the observer would originate within a radius of roughly
2.5 physical kpc. The same percentage of photons after taking into account RT would come from a radius of roughly 2.9 physical kpc.\footnote{Note that the
pixels in the right panel of Figure \ref{rt_only} that give the impression of a diffusion of the photons due to scattering possibly larger than our $\sim 10\%$ estimate,
correspond to pixels with practically zero number of photons.}

So far we have been ignoring the GP absorption. When adding  this absorption  we consider two distinct cases. 
In the first case we include the red damping wing of the GP absorption, and in the second case we set it equal to zero. The latter best-case
scenario is what would happen if for example the emitter was inside the HII region of a very bright quasar.
The spectrum obtained in the first case is shown with the dotted line in the left
panel of Figure \ref{rt_only}, whereas the spectrum in the second case is shown with the short dashed line.
An image of the emitter as would  appear on earth  with the GP absorption {\it and} the damping wing
included is shown in the right panel of Figure \ref{rt_only}. 

Not surprisingly, when the damping wing  is not taken into
account the spectrum  is identical with that before the GP absorption with the difference that all  flux blueward of the \lya resonance is missing.
When including the damping wing the maximum flux is suppressed by roughly a factor of 61.7 with respect to the maximum flux without it.
This line is still quite wide, with a width of approximately 1370 km/s at a flux level of $10^{-21}$ ergs s$^{-1}$ cm$^{-2}$ $\AA^{-1}$, and
a FWHM roughly $620$ km/s. 

Lastly, these results have converged with respect to both the number of photons and the radius where the detailed RT stops (and beyond which the GP absorption
is added). More specifically, we find that the number of photons required for the initial (no RT) emission to converge ($3 \times 10^{5}$) 
is enough for the with RT and GP absorption results.  And,  the results also converge if a 10 physical kpc radius is used for the detailed RT and
beyond that the GP absorption is added.
Convergence has been checked also with respect to the minimum cell initial luminosity considered. We find that 
the results  converge if the minimum luminosity discussed in \S \ref{intrinsic_emission} in the context of initial emission convergence is used.

\subsection{Some additional physics considerations}
\label{phys_cons}
Here we discuss the importance of collisions while the photons are propagating, as well as 
the possible role of dust (currently not taken into account).
\subsubsection{Collisions}
\label{sec:collisions}
While the photons are undergoing scattering, 
collisions should be considered in the following three contexts: (i) collisional redistribution within the $n=2$ state; if for example
a collision makes the atom go from the $2P_{3/2}$ to the $2S_{1/2}$, then the \lya photon is destroyed through a 2 photon decay of the
$2S_{1/2}$ state. If instead the collision takes it to the $2P_{1/2}$, the scattering phase function will be different and hence it is relevant
in either case to see how probable the collisional redistribution is (ii) collisional de-excitation of the $n=2$, in which case
the
photon is lost (iii) collisional broadening of the line, which could cause non-coherence in the rest frame of the atom. 
The RT code can take all these processes into account, but here we develop some intuition as to their importance. 
In fact, since, as will be shown, these processes are in practice negligible, the corresponding calculations in the RT code were switched off when 
producing the results presented in this study. 

Referring to cases (i) and (ii), the largest collisional cross sections are for momentum changing transitions  \cite[$\Delta L=\pm1$; e.g.,][]{osterbrock89}.
As discussed already, both collisions with electrons and protons are relevant, but protons are more significant in case (i), whereas electrons are more
significant in case (ii). We have already calculated the probability per scattering that the $2P \rightarrow 2S$ 
transition of case (i) happens (see equation (\ref{collisions_prob})).
The maximum value of this probability for the conditions of the simulation cells is roughly 
$10^{-10}$ (assuming $T_{4}=1, n_{p}=10^{2}$ cm$^{-3}$, with the latter being of the order of the maximum proton 
number density of cells in simulations. The temperature
dependence is so weak that it does not really matter what temperature one assumes, for order of magnitude estimates). 
So, unless a photon undergoes $10^{10}$ scatterings, collisions of the type (i) should not matter. The cells that are relevant for this
are optically thick cells where the photons scatter repeatedly. Since as we saw $N_{sc} \simeq \tau_{0}$ 
and none of the simulation cells has $\tau_{0}$ larger than a few times $10^{9}$, collisions should not have a significant impact. 
Note that for most
cells the number of scatterings for which collisions may start to matter is orders of magnitude higher than $10^{10}$ (i.e., what 
is described above is the worst case scenario as far as the effect of collisions is concerned since it assumes the {\it maximum} 
proton number density, present
in very few cells).
If these collisions do not matter then collisions of type (ii), which have smaller cross sections, should not matter either. 
   
In case (iii), if the atom suffers collisions with other particles while it is emitting, the phase of the emitted radiation can be altered
suddenly. If the phase changes completely randomly at the collision times, then information about the emitting radiation is lost and coherence is
destroyed.  In this case, in the rest frame of the atom, the line profile is Lorentzian but the total width is the natural width plus the frequency of
collisions the atom experiences on average. 
Since the importance of this effect as well is determined by a comparison of the radiative decay time and the
time between collisions (i.e., equation \ref{collisions_prob}), from the above discussion it becomes clear that it is also negligible. 

\subsubsection{Dust}
\label{sec:dust}
Dust absorbs \lya photons.  Thus, one would assume that dust in the presence of scattering that  traps photons, could have
a significant effect, and that this may be true  even if it is present in small amounts, as  is expected to be the case for the  $z\simeq 8$ 
emitter we discuss  (with a metallicity  roughly equal to 0.1 the solar metallicity). Indeed, \cite{charlot_fall91} found that only a tiny fraction of \lya photons escape from
a static, neutral ISM even if there is  a tiny amount of dust present. To include the effect of dust absorption in simulations
we will have to implement a recipe to estimate the amount of dust. 
Even though one can come up  with an observationally motivated recipe (albeit with 
unknown applicability at redshifts as high as 8), we postpone such a treatment
for a future study, since the main focus of the current study is the \lya RT scheme (which nevertheless includes the probability per scattering
that the photon will be absorbed, but this probability is currently set to 0). 

However, the \lya emitter results  we present in this study  should not be  taken as unrealistic, since it is not obvious 
how these results will change if we include the effects of  dust.
More specifically, many starforming galaxies are observed to have
significant \lya luminosities \cite[e.g.,][]{kunth_etal98,pettini_etal00}, and this is usually attributed to the presence of galactic winds in these
systems that allow the \lya photons to escape after much fewer scatterings than in the static medium case. 
These data seem to support the idea that it is the kinematics  of the gas rather than the dust content that is the dominant \lya escape regulator.

Furthermore, \cite{neufeld91}  found that under suitable conditions the effects of dust absorption may actually increase rather than diminish
the observed \lya line strength relative to radiation that suffers little or no scattering. This would happen for example in a multiphase medium consisting
of dusty clumps of neutral hydrogen embedded within a relatively 'transparent' medium. If most of the dust lies in cold neutral clouds then \lya photons, not
being able to penetrate those clumps, will not be affected as much by the presence of dust \cite[see also][]{hansen_oh05}. Although 
there is no direct observational
evidence to support this structure for the ISM (i.e., that dust lies preferentially in cold, neutral hydrogen clumps, even though the clumpiness in the
distribution of neutral hydrogen itself seems to be established observationally \cite[see][and references therein]{hansen_oh05}),
such a morphology of the dust and atomic hydrogen distribution
could help account for the lack of strong correlation between dust content -- inferred from metallicity or submillimeter emission -- 
and \lya equivalent width. For example, some dust-rich galaxies have substantially higher \lya escape fraction than less dusty emitters
\citep{kunth_etal98,kunth_etal03}. In addition, \cite{giavalisco_etal96} found that there is no correlation between the \lya equivalent widths and the slope
of the UV continuum, which is a measure of the continuum extinction and hence of dust content.

Another reason why it is not obvious how the results presented here will change if we  take dust into account, 
 is that in the current version of the ART code  molecular hydrogen forms only through the catalytic action of electrons.
 When molecular hydrogen formation on grains is included in the code, 
some of what is currently taken to be neutral atomic hydrogen will transform into  
molecular hydrogen, hence this  effect will  decrease the optical thickness of what currently are the thickest cells.

\section{Summary }
\label{sec:conclusions}
We develop a  \lya RT  code applicable to gasdynamics cosmological simulations.
High resolution,  along with appropriately treated cooling can lead 
to very optically thick environments.
Solving the \lya RT even for one  very thick simulation cell takes a long  time.  Solving it for the whole simulation box,  or a significant 
fraction of it, takes unrealistic time. 
Thus, we develop accelerating schemes to speed up the RT.
We treat the moderately thick cells by skipping the numerous core scatterings which are not associated with any significant spatial
diffusion, and go directly to the scattering that takes the photon outside of the core. We use depth dependent core definitions, and find
that quite  large core values can be used. 
For the very optically thick cells we motivate our treatment from the classical problem of resonant radiation transfer in a semi-infinite slab.
We find that with some modifications, since the simulations have  cubic cells rather than slabs, 
we can use the analytical solution derived by \citet{neufeld90} for the problem of the semi-infinite slab.
With these accelerating methods, along with the parallelization of the code we made the problem of \lya RT in the complex environments of
cosmological simulations tractable and solvable.
Even though our approach assumes  a cell structure for the simulation outputs, as is inherently the case in AMR codes,
the \lya RT code we discuss  is applicable to outputs from all kinds of cosmological simulation codes.
This is true  since 
one can always create an effective mesh by interpolating the values of the various physical parameters. 

We perform a series of tests of the RT code, and then  we apply it to  ART cosmological simulations.
We focus on the  brightest emitter in those simulations at $z\simeq 8$. 
A first interesting result for this emitter pertains to its intrinsic emission region and mechanisms.
The emission region consists of two  smaller regions, each corresponding to one of the two main progenitors that merged to form the
emitter at $z\simeq 8$. Both regions are pretty small, with the larger of the two having a diameter of $2-2.5$ physical kpc.
Furthermore, recombination produced \lya photons is the dominant intrinsic \lya emission mechanism, with collisional excitation and SNR produced
\lya photons being subdominant. 
The intrinsic luminosity of the emitter is $4.8 \times 10^{43}$ ergs/s, whereas the injection spectrum (i.e., initial emission spectrum) 
shows significant velocity structure.
 
After performing the \lya RT, but before adding the GP absorption, the emitter spectrum obtained 
resembles that of a very optically thick static configuration, despite the slight trend  for inward radial
motions. More specifically, we obtain the usual double horn spectrum. This happens because (i) even though there is
some net inward radial motion, there are still significant tangential peculiar velocity components, and (ii) the optical
depth is so high that velocities of order some hundreds km/s will not favor blue versus red photons (i.e., in order to escape,
 both kinds of photons have
to shift off resonance much more than  the shift because of peculiar velocities, thus none of the two kinds of photons is  favored in particular 
because of the existence of bulk motions).
Namely,  the velocity information is in fact lost because of the extremely high optical depth.
The width of the two horns is noticeably high ($\sim 5000$ km/s), but in agreement with what is expected 
for the  high simulation column densities. The size of the emitter increases, since the scatterings disperse the photons on a larger area.
We find that on the plane of the emitter image, a photon on average escapes at a distance 
of about $10 \%$ of the initial (before RT) emitter size from the point it was originally emitted.

We include the GP absorption in two different ways: without and with the red damping wing. In the first case the spectrum is identical
to that when the GP is not included, with the difference that now we get only the red peak (rather than both the red and blue peaks). 
This case would correspond to the situation where the \lya emitter lies within the HII region of a very bright quasar.
In the second case, where the damping wing is taken into account, the red peak is also affected.
Its  maximum flux is  suppressed compared to when no damping wing is used by roughly a factor of 61.7. The resulting line after including the wing
is still quite broad with a velocity width of about $1350$ km/s at a flux level of $10^{-21}$ergs s$^{-1}$ cm$^{-2}$ $\AA^{-1}$, and a FWHM of
about $620$ km/s.
The line is quite displaced redward from the \lya resonance, and reach a maximum monochromatic flux of $10^{-20.2}$ ergs s$^{-1}$ cm$^{-2}$ $\AA^{-1}$.

Attempting a detailed  comparison with existing observations, or discussing detection prospects for an object  such as
the simulated emitter is beyond the scope of this study. We have studied only one emitter, and this for only one direction of observation since
our main goal was to use it as an application for the \lya RT code. Thus, we do not have a large enough and representative simulation sample yet.
Furthermore, currently the highest redshift where a \lya line has been observed is $\sim 6.6$ \citep{kodaira_etal03}\footnote{The detection of a 
$z=10$ \lya emitting galaxy was recently reported by \citet{pello_etal04} following a color selected  survey for $z>7$ galaxies located behind a well
studied gravitational lens cluster, but the exact nature of this source remains contentious \citep[e.g.,][]{weatherley_etal04}}  and
it is not  known how different the properties of higher redshift emitters are from that of lower redshift ones.
The most recent report at $z=9$ is that of \citet{willis_courbin05}.  This study finds no detections.
The sky area coverage is possibly 
a significant factor contributing to this no detection result. 
Instead, we  content ourselves here with a simple order of magnitude comparison.  
The intrinsic \lya luminosity of our emitter is consistent with luminosities reported in literature. For example, the
highest \lya luminosity of the $z=5.7$ sample of \citet{hu_etal04} is roughly $6 \times 10^{43}$ ergs/s. 
Higher luminosities than those have been inferred for \lya blobs, rather than emitters. For example, the most
luminous blob in the sample of \citet{matsuda_etal04} has a \lya luminosity of $1.1 \times 10^{44}$ ergs/s. 
Most  observed \lya emitters are unresolved and so is expected to be the simulated emitter. 
Reported sizes for the observed objects are in the $\sim$ few kpc range \cite[e.g.,][]{hu_etal02}.
\lya blobs on the other hand are quite more extended,  with sizes $\sim 100$ kpc \citep{matsuda_etal04}.
The widths of the (lower z) observed lines are typically a few hundred km/s, whereas the FWHM of the simulated line is roughly $620$ km/s.
As discussed already, the  velocity width of the ART emitter could be affected by the very high H column densities which will
drop as soon as molecular hydrogen formation on dust grains is taken into account. 
In terms of the detectability, if one adopts the present day
limit of ground based detections of  $\sim 10^{-18}$ ergs s$^{-1}$ cm$^{-2}$ $\AA^{-1}$, clearly our simulated emitter
would be orders of magnitude fainter. If the emitter  is embedded within the HII region of a bright quasar, in which case the red damping wing
will be suppressed, the brightness is marginally below the sensitivity of current ground based instruments.
Note, however, that the prospects of detection  will be much better for JWST which is expected to be able to detect $\sim 400$ times fainter objects than currently
studied with ground based infrared telescopes.

\acknowledgements 
I am grateful to   N. Y. Gnedin and A.V. Kravtsov 
for many useful discussions and guidance,  for comments on the manuscript, and for allowing me to use their simulations.
I would like to thank P. Jonsson, D. Neufeld, J. Rhoads, Y. Tanigucchi, and Z. Zheng  for fast and
comprehensive responses to  my questions.  This work benefited greatly from my interaction
with J. Carlstrom, A. Konigl, and A.V. Olinto, and was supported by the National Science Foundation (NSF) under grants
ASTR 02-06216 and ASTR 02-39759, by NASA through grants NAG5-13274 and NAG5-12326, and by  the Kavli Institute for Cosmological Physics
at the University of Chicago.
The author also acknowledges support through an award from the Onassis Foundation. 
The simulations  discussed  were performed on Linux Clusters and IBM690 arrays at the National
Center for Supercomputer Applications and the San Diego Supercomputer Center under the National Partnership for
Advanced Computational Infrastructure grant $\#$MCA03S023.
This work was presented as part of a dissertation to the Department of Astronomy and Astrophysics, The University of Chicago, in 
partial fulfillment of the requirements for the Ph.D. degree.
 
\bibliography{tasitsiomi_lyman}
\end{document}